\documentclass[letterpaper,11pt]{article}
\def\OMIT#1{}
\newcommand{\leng}[1]{|{#1}|}
\catcode`\@=11
\def\newexample#1{\@ifnextchar[{\@oexm{#1}}{\@nexm{#1}}}

\def\@nexm#1#2{%
\@ifnextchar[{\@xnexm{#1}{#2}}{\@ynexm{#1}{#2}}}

\def\@xnexm#1#2[#3]{\expandafter\@ifdefinable\csname #1\endcsname
{\@definecounter{#1}\@addtoreset{#1}{#3}%
\expandafter\xdef\csname the#1\endcsname{\expandafter\noexpand
  \csname the#3\endcsname \@exmcountersep \@exmcounter{#1}}%
\global\@namedef{#1}{\@exm{#1}{#2}}\global\@namedef{end#1}{\@endexample}}}

\def\@ynexm#1#2{\expandafter\@ifdefinable\csname #1\endcsname
{\@definecounter{#1}%
\expandafter\xdef\csname the#1\endcsname{\@exmcounter{#1}}%
\global\@namedef{#1}{\@exm{#1}{#2}}\global\@namedef{end#1}{\@endexample}}}

\def\@oexm#1[#2]#3{\expandafter\@ifdefinable\csname #1\endcsname
  {\global\@namedef{the#1}{\@nameuse{the#2}}%
\global\@namedef{#1}{\@exm{#2}{#3}}%
\global\@namedef{end#1}{\@endexample}}}

\def\@exm#1#2{\refstepcounter
    {#1}\@ifnextchar[{\@yexm{#1}{#2}}{\@xexm{#1}{#2}}}

\def\@xexm#1#2{\@beginexample{#2}{\csname the#1\endcsname}\ignorespaces}
\def\@yexm#1#2[#3]{\@opargbeginexample{#2}{\csname
       the#1\endcsname}{#3}\ignorespaces}

\def\@exmcounter#1{\noexpand\arabic{#1}}
\def\@exmcountersep{.}
\def\@beginexample#1#2{\trivlist \item[\hskip 
\labelsep{\bf #1\ #2:}]}
\def\@opargbeginexample#1#2#3{\trivlist
      \item[\hskip \labelsep{\bf #1\ #2\ }#3{\bf :}]}
\def\@endexample{\endtrivlist}

\catcode`\@=12

\newtheorem{lemma}{{\bf Lemma}}[section]
\newtheorem{thm}{{\bf Theorem}}[section]
\newtheorem{df}{{\bf Definition}}[section]
\newtheorem{cor}{{\bf Corollary}}[section]
\newtheorem{prop}{{\bf Proposition}}[section]

\newcommand{\BQED}{\hfill \hbox{\rule{8pt}{8pt}}}
\newcommand{\cmb}[2]{{#1\choose#2}}

\newenvironment{namelist}[1]{%
\begin{list}{}
  { 
	\settowidth{\labelwidth}{#1}
	\setlength{\leftmargin}{1.1\labelwidth}}
  \setlength{\itemsep}{0cm}
}{% 
\end{list}}

\catcode`\@=11
\renewcommand{\@biblabel}[1]{\hspace*{\fill}[#1]}
\catcode`\@=12
\begin{document}
%
%\vspace*{0.7cm}
%
\begin{center}
{\LARGE {\bf Weighted Random Popular Matchings}}\bigskip\\
\begin{tabular}{ccc}
{\sc Toshiya Itoh} & & {\sc Osamu Watanabe}\\
{\sf titoh@dac.gsic.titech.ac.jp} & & 
{\sf watanabe@is.titech.ac.jp}\smallskip\\
%{\sf Global Scientific Information and Computing Center} & & 
{\sf Global Sci. Inform. and Comput. Center} & & 
%{\sf Department of Mathematical and Computing Systems}\\
{\sf Dept. of Math. and Comput. Sys.}\\
{\sf Tokyo Institute of Technology} & & {\sf Tokyo Institute of Technology}\\
%{\sf GSIC} & & 
%{\sf Global Sci. and Inform. Comput.}\\
%{\large {\sf Interdisciplinary Graduate School of Science and Engineering}}\\
%{\sf Tokyo Institute of Technology} & & 
%{\sf Tokyo Institute of Technology}\\
{\sf Meguro-ku, Tokyo 152-8550, Japan} & & {\sf Meguro-ku, Tokyo 152-8552, Japan}
%{\sf Tel: +81-3-5734-3651}~~~~~{\sf titoh@dac.gsic.titech.ac.jp}
%
\end{tabular}
\end{center}
{\bf Abstract:}
For a set $A$ of $n$ applicants and a set $I$ of $m$ items,
we consider a problem of computing~a~matching of applicants to items,
i.e., a function ${\cal M}$ mapping $A$ to $I$;
here we assume that each applicant $x \in A$ provides 
a {\it preference list\/} on items in $I$.
We say that an applicant $x \in A$ {\it prefers\/} 
an~item~$p$~than~an~item~$q$~if $p$ is located at 
a higher position than $q$ in its preference list,
and we say that $x$ {\em prefers\/} a matching~${\cal M}$~over a matching ${\cal M}'$
if $x$ prefers ${\cal M}(x)$ over ${\cal M}'(x)$.
For a given matching problem $A$, $I$, and preference lists,~we 
say that ${\cal M}$ is {\it more popular\/} than ${\cal M}'$
if the number of applicants preferring
${\cal M}$ over ${\cal M}'$ is larger than that 
of applicants preferring ${\cal M}'$ over ${\cal M}$,
and ${\cal M}$ is called a {\it  popular matching\/}
if there is no other matching that is more popular than ${\cal M}$.
Here we consider the situation that
$A$ is partitioned into $A_{1},A_{2},\ldots,A_{k}$,~and that
each $A_{i}$ is assigned a weight $w_{i}>0$ such that $w_{1}>w_{2}>\cdots>w_{k}>0$.
For~such~a~matching~problem, 
we say that ${\cal M}$ is {\it more popular\/} than ${\cal M}'$
if the total weight of applicants preferring ${\cal M}$ over ${\cal M}'$
is larger than that of applicants preferring ${\cal M}'$ over ${\cal M}$,
and we call ${\cal M}$
an {\it $k$-weighted popular matching\/}
if there is no other matching that is more popular than ${\cal M}$. 
Mahdian [In Proc. of the~7th~ACM~Conference~on Electronic Commerce, 2006]
showed that if $m > 1.42 n$,
then a random instance~of~the~(nonweighted) matching problem
has a popular matching with high probability.
In this paper,
we analyze the 2-weighted matching problem,
and we show that (lower bound)
if $m/n^{4/3}=o(1)$,
then a random instance of the 2-weighted matching problem with $w_{1} \geq 2w_{2}$
has a 2-weighted popular matching with probability $o(1)$;~and 
(upper bound) if $n^{4/3}/m = o(1)$, then a random instance of 
the 2-weighted matching problem with $w_{1} \geq 2w_{2}$~has 
a 2-weighted popular matching with probability $1-o(1)$.\bigskip\\
{\bf Key Words:} %Popular Matchings, 
Random Popular Matchings, 
Weighted Popular Matchings, 
Well-Formed Matchings.
%
%\newpage
%
%===========================================
\section{Introduction} \label{introduction}
%===========================================
%
%==========================================
%\subsection{Background} \label{background}
%==========================================
%
For a set $A$ of $n$ applicants and a set $I$ of $m$ items,
we consider the problem of computing~a~certain~matching of applicants to items,
i.e., a function ${\cal M}$ mapping $A$ to $I$.
Here we assume that each applicant~$x \in A$~pro\-vides its 
{\it preference list\/} defined on a subset $J_{x} \subseteq I$.
A preference list $\vec{\ell}_{x}$ of each applicant $x$ may contain~ties 
among the items and it ranks subsets $J_{x}^{h}$'s of $J_{x}$; 
that is, $J_{x}$ is partitioned into $J_{x}^{1},J_{x}^{2},\ldots,J_{x}^{d}$,
where $J_{x}^{h}$~is a set of the $h^{th}$ preferred items.
We say that an applicant $x$ {\it prefers\/} $p \in J_{x}$ than $q \in J_{x}$
if $p \in J_{x}^{i}$ and $q \in J_{x}^{h}$ for $i < h$. 
For any matchings ${\cal M}$ and ${\cal M}'$, we say that an applicant 
$x$ {\it prefers\/} ${\cal M}$ over ${\cal M}'$ 
if the applicant $x$ prefers ${\cal M}(x)$ over ${\cal M}'(x)$, 
and we say that ${\cal M}$ is {\it more popular\/} than ${\cal M}'$
if the total number of applicants
preferring ${\cal M}$ over ${\cal M}'$ is larger than that of 
applicants preferring 
${\cal M}'$~over~${\cal M}$.~${\cal M}$~is~called~a~{\it popular~match\-ing\/} \cite{G}
if there is no other matching that is more popular than ${\cal M}$.
The {\em popular matching problem\/} 
is to compute this popular matching for given $A$, $I$, and preference lists.
This problem has applications~in~the real world,
e.g., mail-based DVD rental systems such as NetFlix \cite{ACKM}.

Here we consider the (general) situation
that the set $A$ of applicants is partitioned into
several categories $A_{1},A_{2},\ldots,A_{k}$,
and that each category $A_{i}$ is assigned a weight $w_{i}>0$
such that $w_{1}>w_{2}>\cdots >w_{k}$. 
This setting can be regarded as a case where 
the applicants in $A_{1}$ are platinum members, the applicants 
in $A_{2}$ are gold members,
the applicants in $A_{3}$ are silver members,
the applicants in $A_{4}$ are regular members, etc. 
In a way similar to the above,
we define the {\it $k$-weighted popular matching problem\/} \cite{J.M}, 
where the goal is to compute a popular matching ${\cal M}$
in the sense that for any other matching ${\cal M}'$, 
the total weight of applicants preferring ${\cal M}$
is larger than that of applicants preferring ${\cal M}'$. 
Notice that the original popular matching problem, 
which we will call the {\it single category} popular matching problem, 
is the 1-weighted popular matching problem. 

We say that a preference list $\vec{\ell}_{x}$ of an applicant $x$ is 
{\it complete\/} if $J_{x}=I$, 
that is, $x$ shows its preferences on all items,
and a $k$-weighted popular matching problem $(A,I,\{\vec{\ell}_{x}\}_{x \in A})$
is called {\it complete\/} if $\vec{\ell}_x$ is complete for every applicant $x \in A$. 
We also say that
a preference list $\vec{\ell}_{x}$  of an applicant $x$ is {\it strict\/} if 
$\leng{J_{x}^{h}}=1$ for each $h$, 
that is,
$x$ prefers each item in $J_{x}$ differently,
and a $k$-weighted popular matching problem is called {\it strict\/}
if $\vec{\ell}_{x}$ is strict for every applicant $x \in A$.
%
%========================================
\subsection{Known Results} \label{known}
%========================================
%
For the strict single category popular matching problem,
Abraham, et al.\ \cite{AIKM}
presented a deterministic $O(n+m)$ time algorithm
that outputs a popular matching if it exists;
they also showed, 
for the single category popular matching problem with ties,
a deterministic $O(\sqrt{n}m)$ time algorithm. To~derive~these~al\-gorithms, 
Abraham, et al.\ introduced the notions of $f$-items
(the first items) and $s$-items (the second items), and 
characterized popular matchings by $f$-items and $s$-items.
Mestre \cite{J.M} generalized those results 
to the $k$-weighted popular matching problem,
and he showed a deterministic $O(n+m)$ time algorithm for the strict case, 
%TI: 追加して見ました．くどい？
where it outputs a $k$-weighted popular matching if any, 
and a deterministic $O(\min(k\sqrt{n},n)m)$ time algorithm for the case with ties.

In general, some instances of the complete and strict single category 
popular matching problem do not have a popular matching. Answering to a question of
when a random instance of the complete and strict 
single category popular matching problem
has a popular matching,
Mahdian \cite{M.M} showed that if $m > 1.42n$,
then a random instance of the popular matching problem has 
a popular matching with probability $1-o(1)$;
he also showed that
if $m<1.42 n$,
then a random instance of the popular matching problem 
has a popular matching with probability $o(1)$.
%
%=====================================
\subsection{Main Results} \label{main}
%=====================================
%
In this paper,
we consider the complete and strict 2-weighted popular matching problem,
and investigate when a random instance of the complete and strict 
2-weighted popular matching problem has a 2-weighted popular matching.
Our results are summarized as follows.\medskip\\
{\bf Theorem \ref{thm-lower}:} 
%Let $m = \beta n$. 
If $m/n^{4/3}=o(1)$,
then a random instance of the complete and strict $2$-weighted popular 
matching problem with $w_{1}\geq 2w_{2}$ has 
a $2$-weighted popular matching with probability $o(1)$.\medskip\\
{\bf Theorem \ref{thm-upper}:} 
%Let $m = \beta n$. 
If $n^{4/3}/m = o(1)$,
then a random instance of the complete and strict $2$-weighted popular matching problem
with $w_{1}\geq 2w_{2}$ has 
a $2$-weighted popular matching with probability $1 - o(1)$.\medskip

For an instance of the single category popular matching problem,
it suffices to consider only a set $F$ of $f$-items and a set $S$ of $s$-items 
\cite{M.M}. 
For an instance of the 2-weighted popular matching problem, however,
we need to {\em separately\/} consider
$f_{1}$-items, $s_{1}$-items, $f_{2}$-items, and $s_{2}$-items; 
let $F_1$, $S_1$, $F_2$, and $S_2$ denote these item sets.
Some careful analysis is necessary,
in particular, because in general, 
we may have the situation $S_{1}\cap F_{2}\neq \emptyset$,
which makes our probabilistic analysis much harder than (and quite different from) 
the single category case. 
%
%=============================================
\section{Preliminaries} \label{preliminaries}
%=============================================
%
In the rest of this paper,
we consider the complete and strict 2-weighted popular matching problem.
Let~$A$ be the set of $n$ applicants and $I$ be the set of $m$ items.
%For some $\beta\geq 1$, let $m=\beta n$.
We assume that $A$ is partitioned into $A_{1}$ and $A_{2}$, 
and we refer to $A_{1}$ (resp. $A_{2}$) as
the first (resp. the second) category. 
For any constant $0 < \delta < 1$,
we also assume that
$\leng{A_{1}}=\delta\leng{A}=\delta n$ and 
$\leng{A_{2}}=(1-\delta)\leng{A}=(1-\delta) n$.
Let $w_{1}> w_{2}>0$ 
be weights of the first category $A_{1}$ and the second category $A_{2}$, respectively.

We define $f$-items and $s$-items \cite{AIKM,J.M} as follows:
For each applicant $x \in A_{1}$,
let $f_{1}(x)$ be the most preferred item in its preference list $\vec{\ell}_{x}$,
and we call it an {\it $f_1$-item of $x$\/}.
We use $F_{1}$ to denote the set~of~all $f_{1}$-items of applicants $x \in A_{1}$. 
For each applicant $x \in A_{1}$,
let $s_{1}(x)$ be the most preferred item in its preference list $\vec{\ell}_{x}$
that is not in $F_{1}$,
and we use $S_{1}$ to denote the set of all $s_{1}$-items of applicants $x \in A_{1}$. 
Similarly,
for each applicant $y \in A_{2}$, 
let $f_{2}(y)$ and $s_2(y)$ be
the most preferred item in its preference list $\vec{\ell}_{y}$ 
that is not in $F_1$ and not in $F_1\cup F_2$, respectively,
where we use $F_2$ and $S_2$ to denote the set of all $f_2$-items 
and $s_2$-items, respectively. 
From this definition, we have that 
$F_{1}\cap S_{1}=\emptyset$, $F_{1}\cap F_{2}=\emptyset$, and  
$F_{2}\cap S_{2}=\emptyset$;
on the other hand,
we may have that
$S_{1}\cap F_{2}\neq \emptyset$ or $S_{1}\cap S_{2}\neq \emptyset$.

For characterizing the existence of $k$-weighted popular matching,
Mestre \cite{J.M} defined the notion~of~``well-formed matching,''
which generalizes well-formed matching 
for the single category popular matching~prob\-lem \cite{AIKM}. 
We recall this characterization here.
Below we consider any instance $(A,I,\{\vec{\ell}_{x}\}_{x \in A})$ of 
the strict (not necessarily complete) $2$-weighted popular matching problem.
\begin{df} \label{df-well-formed-1}
A matching ${\cal M}$ is {\sf well-formed}
if by ${\cal M}$
{\rm (1)} each $x \in A_{1}$ is matched to $f_{1}(x)$ or $s_{1}(x);$
{\rm (2)} each each $y \in A_{2}$ is matched to $f_{2}(y)$ or $s_{2}(y);$
{\rm (3)} each $p\in F_{1}$ is matched
to some $x \in A_{1}$ such that $p=f_1(x);$
and {\rm (4)} each $q\in F_{2}$ is matched
to some $y \in A_{2}$ such that $q=f_2(y)$.
\end{df}
Mestre \cite{J.M} showed that
the existence of a 2-weighted popular matching is almost equivalent
to that~of~a~well-formed matching.
Precisely,
he proved the following characterization.
\begin{prop}[\cite{J.M}] \label{prop-mestre}
Let $(A,I,\{\vec{\ell}_{x}\}_{x \in A})$ be an instance of the strict 
$2$-weighted popular matching problem. Any $2$-weighted popular matching of 
$(A,I,\{\vec{\ell}_{x}\}_{x \in A})$ 
is a well-formed matching, and if $w_1\ge 2w_2$,
then any well-formed matching of $(A,I,\{\vec{\ell}_{x}\}_{x \in A})$ 
is a $2$-weighted popular matching. 
\end{prop}

Consider an instance $(A,I,\{\vec{\ell}_{x}\}_{x \in A})$ of 
the strict (not necessarily complete)
$2$-weighted popular~match\-ing problem with weights $w_1\ge 2w_2$.
As shown above, the existence of 
a 2-weighted popular matching~is characterized by that of a well-formed matching,
which is determined by the structure of $f_1$-,~$f_2$-,~$s_1$-,~and $s_2$-items. 
Here we introduce a graph $G=(V,E)$ for investigating this structure, 
and~in~the~following~discussion, we will mainly use this graph.
The graph $G=(V,E)$ is defined by a set 
$V=F_1\cup S_1\cup F_2\cup S_2$ of vertices, and
the following set $E$ of edges. 
\[
E =
\{(f_1(x),s_1(x)):x\in A_1\}\cup\{(f_2(y),s_2(y)):y\in A_2\}.
\]
We use $E_1$ and $E_2$ to denote
the sets of edges
defined for applicants in $A_1$ and %applicants in 
$A_2$, respectively, i.e., the former and the latter sets of the above.
In the following,
the graph $G=(V,E)$ defined above is called an 
{\it fs-relation graph\/} for $(A,I,\{\vec{\ell}_{x}\}_{x \in A})$. 
Note that
this fs-relation graph
$G=(V,E)$ consists of $M=\leng{V} \leq m$ vertices and $n=\leng{A}$ edges.
If $e_{1} \in E_{1}$ and $e_{2} \in E_{2}$ are incident to the same vertex $p \in V$,
then we have either
$p \in S_{1} \cap F_{2}$ or $p \in S_{1} \cap S_{2}$.
This situation makes the analysis of the 2-weighted popular matching problem
harder than and different from the one for the single category case.

%For an instance $(A,I,\{\vec{\ell}_{x}\}_{x \in A})$ of the strict 
%2-weighted popular matching problem, 
We now characterize the existence of a well-formed matching as follows.
\begin{lemma} \label{lemma-orientation}
An instance $(A,I,\{\vec{\ell}_{x}\}_{x \in A})$ of 
the strict $2$-weighted popular matching problem
has a well-formed matching
iff its fs-relation graph $G=(V,E)$ has an orientation ${\cal O}$ on edges
such that {\rm (a)}~each~$p \in V$ has at most one incoming edge in $E_{1} \cup E_{2};$
{\rm (b)} each $p \in F_{1}$ has one incoming edge in $E_{1};$
and {\rm (c)} each $q \in F_{2}$ has one incoming edge in $E_{2}$.
\end{lemma}
{\bf Proof:}
Consider any instance $(A,I,\{\vec{\ell}_{x}\}_{x \in A})$
of the strict 2-weighted popular matching problem,
where $A=A_1\cup A_2$,
and let $G=(V,E)$ be its fs-relation graph.

First assume that
this instance has a well-formed matching ${\cal M}$.
Define an orientation ${\cal O}$ on edges~of~the graph $G=(V,E)$ as follows:
For each applicant $a \in A_{i}$, orient an edge 
$e_{a}=(f_{i}(a),s_{i}(a)) \in E_{i}$ 
toward ${\cal M}(a)$. Since ${\cal M}$ is a matching 
between $A$ and $I$, we 
have that each $p \in V$ has at most one incoming edge. 
From the condition (3) of Definition \ref{df-well-formed-1}, 
it follows that each $p \in F_{1}$ has one incoming 
edge in $E_{1}$,~and~from the condition (4) of Definition \ref{df-well-formed-1}, 
it follows that each $q \in F_{2}$ has one 
incoming~edge~in~$E_{2}$.~Thus~the~ori\-entation ${\cal O}$ on edges of $G=(V,E)$ 
satisfies the conditions (a), (b), and (c). 

Assume that the graph $G=(V,E)$ has an orientation ${\cal O}$ 
on edges satisfying the conditions (a),~(b),~and (c). 
Then we define a matching ${\cal M}$ as follows:
For each $x \in A_{1}$, its $f_{1}$-item $f_{1}(x)$ (resp. $s_{1}$-item 
$s_{1}(x)$)~is matched to $x$ if ${\cal O}$ 
orients the edge $e_{x}=(f_{1}(x),s_{1}(x))\in E_{1}$ 
by $f_{1}(x) \leftarrow s_{1}(x)$ (resp. $f_{1}(x) \rightarrow s_{1}(x)$),~and 
for each $y \in A_{2}$, its $f_{2}$-item $f_{2}(y)$ 
(resp. $s_{2}$-item $s_{2}(y)$) is matched to $y$ if ${\cal O}$ orients the edge 
$e_{y}=(f_{2}(y),s_{2}(y))\in E_{2}$ by $f_{2}(y) \leftarrow s_{2}(y)$ 
(resp. $f_{2}(y) \rightarrow s_{2}(y)$). 
From the condition (a) of the orientation ${\cal O}$, it is immediate to see 
that ${\cal M}$ is a matching for $(A,I,\{\vec{\ell}_{x}\}_{x \in A})$. 
From the definition of the graph $G=(V,E)$, we have that ${\cal M}$ satisfies 
the conditions (1) and (2) of Definition \ref{df-well-formed-1}. 
The condition (b) of the orientation ${\cal O}$ implies that 
each $p \in F_{1}$ is matched to $x \in A_{1}$ by ${\cal M}$, 
where $f_{1}(x)=p$, and the condition (c) of the orientation 
${\cal O}$ guarantees that 
each $q \in F_{2}$ is matched to $y \in A_{2}$ by ${\cal M}$, where 
$f_{2}(y)=q$. Thus the matching ${\cal M}$ for 
$(A,I,\{\vec{\ell}_{x}\}_{x\in A})$ satisfies the conditions 
(1), (2), (3), and (4) of 
Definition \ref{df-well-formed-1}. \BQED
%
%======================================================================
\section{Characterization for the 2-Weighted Popular Matching Problem}
\label{sec-characterization}
%======================================================================
%
In this section,
we present necessary and sufficient conditions
for an instance of the strict 2-weighted popular matching problem
to have a 2-weighted popular matching.
For an instance $(A, I, \{\vec{\ell}_{x}\}_{x \in A})$~of~the 
strict 2-weighted popular matching problem,
let $G=(V,E)$ be its fs-relation graph, and consider~the~sub\-graphs 
$G_{1}$, $G_{2}$, and $G_{3}$ of the 
graph $G=(V,E)$ as in Figure \ref{fig-bad-component}. 
\begin{figure}[ht]
\begin{center}
\begin{tabular}{ccc}
\hspace*{-1.25cm}
%
%WinTpicVersion3.08
\unitlength 0.1in
\begin{picture}( 23.7500, 14.0000)(  1.2500,-18.0000)
% LINE 1 0 3 0
% 2 1000 1000 1000 1800
% 
\special{pn 13}%
\special{pa 1000 1000}%
\special{pa 1000 1800}%
\special{fp}%
% LINE 1 0 3 0
% 2 2200 1000 2200 1800
% 
\special{pn 13}%
\special{pa 2200 1000}%
\special{pa 2200 1800}%
\special{fp}%
% STR 2 0 3 0
% 3 1000 900 1000 1000 5 0
% $\bullet$
\put(10.0000,-10.0000){\makebox(0,0){$\bullet$}}%
% STR 2 0 3 0
% 3 2200 900 2200 1000 5 0
% $\bullet$
\put(22.0000,-10.0000){\makebox(0,0){$\bullet$}}%
% STR 2 0 3 0
% 3 1000 1300 1000 1400 5 0
% $\bullet$
\put(10.0000,-14.0000){\makebox(0,0){$\bullet$}}%
% STR 2 0 3 0
% 3 2200 1300 2200 1400 5 0
% $\bullet$
\put(22.0000,-14.0000){\makebox(0,0){$\bullet$}}%
% STR 2 0 3 0
% 3 2200 1700 2200 1800 5 0
% $\bullet$
\put(22.0000,-18.0000){\makebox(0,0){$\bullet$}}%
% STR 2 0 3 0
% 3 1000 1700 1000 1800 5 0
% $\bullet$
\put(10.0000,-18.0000){\makebox(0,0){$\bullet$}}%
% CIRCLE 2 1 3 0
% 4 1600 1000 2200 1000 2200 1000 1000 1000
% 
\special{pn 8}%
\special{ar 1600 1000 600 600  3.1415927 3.2415927}%
\special{ar 1600 1000 600 600  3.3015927 3.4015927}%
\special{ar 1600 1000 600 600  3.4615927 3.5615927}%
\special{ar 1600 1000 600 600  3.6215927 3.7215927}%
\special{ar 1600 1000 600 600  3.7815927 3.8815927}%
\special{ar 1600 1000 600 600  3.9415927 4.0415927}%
\special{ar 1600 1000 600 600  4.1015927 4.2015927}%
\special{ar 1600 1000 600 600  4.2615927 4.3615927}%
\special{ar 1600 1000 600 600  4.4215927 4.5215927}%
\special{ar 1600 1000 600 600  4.5815927 4.6815927}%
\special{ar 1600 1000 600 600  4.7415927 4.8415927}%
\special{ar 1600 1000 600 600  4.9015927 5.0015927}%
\special{ar 1600 1000 600 600  5.0615927 5.1615927}%
\special{ar 1600 1000 600 600  5.2215927 5.3215927}%
\special{ar 1600 1000 600 600  5.3815927 5.4815927}%
\special{ar 1600 1000 600 600  5.5415927 5.6415927}%
\special{ar 1600 1000 600 600  5.7015927 5.8015927}%
\special{ar 1600 1000 600 600  5.8615927 5.9615927}%
\special{ar 1600 1000 600 600  6.0215927 6.1215927}%
\special{ar 1600 1000 600 600  6.1815927 6.2815927}%
% STR 2 0 3 0
% 3 1350 1700 1350 1800 5 0
% $v_{i_{1}} \in S_{2}$
\put(13.5000,-18.0000){\makebox(0,0){$v_{i_{1}} \in S_{2}$}}%
% STR 2 0 3 0
% 3 1520 1300 1520 1400 5 0
% $v_{i_{2}} \in S_{1}\cap F_{2}$
\put(15.2000,-14.0000){\makebox(0,0){$v_{i_{2}} \in S_{1}\cap F_{2}$}}%
% STR 2 0 3 0
% 3 1350 900 1350 1000 5 0
% $v_{i_{3}} \in F_{1}$
\put(13.5000,-10.0000){\makebox(0,0){$v_{i_{3}} \in F_{1}$}}%
% STR 2 0 3 0
% 3 2550 1700 2550 1800 5 0
% $v_{i_{k}} \in S_{2}$
\put(25.5000,-18.0000){\makebox(0,0){$v_{i_{k}} \in S_{2}$}}%
% STR 2 0 3 0
% 3 2800 1300 2800 1400 5 0
% $v_{i_{k-1}} \in S_{1}\cap F_{2}$
\put(28.0000,-14.0000){\makebox(0,0){$v_{i_{k-1}} \in S_{1}\cap F_{2}$}}%
% STR 2 0 3 0
% 3 2620 900 2620 1000 5 0
% $v_{i_{k-2}} \in F_{1}$
\put(26.2000,-10.0000){\makebox(0,0){$v_{i_{k-2}} \in F_{1}$}}%
% STR 2 0 3 0
% 3 600 1100 600 1200 5 0
% $E_{1}$
\put(6.0000,-12.0000){\makebox(0,0){$E_{1}$}}%
% STR 2 0 3 0
% 3 2600 1100 2600 1200 5 0
% $E_{1}$
\put(26.0000,-12.0000){\makebox(0,0){$E_{1}$}}%
% STR 2 0 3 0
% 3 600 1500 600 1600 5 0
% $E_{2}$
\put(6.0000,-16.0000){\makebox(0,0){$E_{2}$}}%
% STR 2 0 3 0
% 3 2600 1500 2600 1600 5 0
% $E_{2}$
\put(26.0000,-16.0000){\makebox(0,0){$E_{2}$}}%
% VECTOR 2 0 3 0
% 2 700 1200 1000 1200
% 
\special{pn 8}%
\special{pa 700 1200}%
\special{pa 1000 1200}%
\special{fp}%
\special{sh 1}%
\special{pa 1000 1200}%
\special{pa 934 1180}%
\special{pa 948 1200}%
\special{pa 934 1220}%
\special{pa 1000 1200}%
\special{fp}%
% VECTOR 2 0 3 0
% 2 700 1600 1000 1600
% 
\special{pn 8}%
\special{pa 700 1600}%
\special{pa 1000 1600}%
\special{fp}%
\special{sh 1}%
\special{pa 1000 1600}%
\special{pa 934 1580}%
\special{pa 948 1600}%
\special{pa 934 1620}%
\special{pa 1000 1600}%
\special{fp}%
% VECTOR 2 0 3 0
% 2 2500 1200 2200 1200
% 
\special{pn 8}%
\special{pa 2500 1200}%
\special{pa 2200 1200}%
\special{fp}%
\special{sh 1}%
\special{pa 2200 1200}%
\special{pa 2268 1220}%
\special{pa 2254 1200}%
\special{pa 2268 1180}%
\special{pa 2200 1200}%
\special{fp}%
% VECTOR 2 0 3 0
% 2 2500 1600 2200 1600
% 
\special{pn 8}%
\special{pa 2500 1600}%
\special{pa 2200 1600}%
\special{fp}%
\special{sh 1}%
\special{pa 2200 1600}%
\special{pa 2268 1620}%
\special{pa 2254 1600}%
\special{pa 2268 1580}%
\special{pa 2200 1600}%
\special{fp}%
\end{picture}%
&  
\hspace*{1.0cm}
%
%WinTpicVersion3.08
\unitlength 0.1in
\begin{picture}( 13.4000, 20.0000)(  0.6000,-26.0000)
% CIRCLE 1 0 3 0
% 4 1000 1000 1400 1000 1400 980 1400 980
% 
\special{pn 13}%
\special{ar 1000 1000 400 400  0.0000000 6.2831853}%
% STR 2 0 3 0
% 3 1000 1300 1000 1400 5 0
% $\bullet$
\put(10.0000,-14.0000){\makebox(0,0){$\bullet$}}%
% LINE 1 2 3 0
% 2 1000 1400 1000 1800
% 
\special{pn 13}%
\special{pa 1000 1400}%
\special{pa 1000 1800}%
\special{dt 0.045}%
% STR 2 0 3 0
% 3 1000 1690 1000 1790 5 0
% $\bullet$
\put(10.0000,-17.9000){\makebox(0,0){$\bullet$}}%
% LINE 1 0 3 0
% 2 1000 1800 1000 2200
% 
\special{pn 13}%
\special{pa 1000 1800}%
\special{pa 1000 2200}%
\special{fp}%
% STR 2 0 3 0
% 3 1000 2100 1000 2200 5 0
% $\bullet$
\put(10.0000,-22.0000){\makebox(0,0){$\bullet$}}%
% LINE 1 0 3 0
% 2 1000 2200 1000 2600
% 
\special{pn 13}%
\special{pa 1000 2200}%
\special{pa 1000 2600}%
\special{fp}%
% STR 2 0 3 0
% 3 1000 2500 1000 2600 5 0
% $\bullet$
\put(10.0000,-26.0000){\makebox(0,0){$\bullet$}}%
% STR 2 0 3 0
% 3 500 900 500 1000 5 0
% $C$
\put(5.0000,-10.0000){\makebox(0,0){$C$}}%
% STR 2 0 3 0
% 3 1150 1400 1150 1500 5 0
% $v_{i_{k}}$
\put(11.5000,-15.0000){\makebox(0,0){$v_{i_{k}}$}}%
% STR 2 0 3 0
% 3 1150 1700 1150 1800 5 0
% $v_{i_{3}}$
\put(11.5000,-18.0000){\makebox(0,0){$v_{i_{3}}$}}%
% STR 2 0 3 0
% 3 1500 2100 1500 2200 5 0
% $v_{i_{2}} \in S_{1} \cap F_{2}$
\put(15.0000,-22.0000){\makebox(0,0){$v_{i_{2}} \in S_{1} \cap F_{2}$}}%
% STR 2 0 3 0
% 3 1150 2500 1150 2600 5 0
% $v_{i_{1}}$
\put(11.5000,-26.0000){\makebox(0,0){$v_{i_{1}}$}}%
% STR 2 0 3 0
% 3 600 1900 600 2000 5 0
% $E_{1}$
\put(6.0000,-20.0000){\makebox(0,0){$E_{1}$}}%
% VECTOR 2 0 3 0
% 2 700 2000 1000 2000
% 
\special{pn 8}%
\special{pa 700 2000}%
\special{pa 1000 2000}%
\special{fp}%
\special{sh 1}%
\special{pa 1000 2000}%
\special{pa 934 1980}%
\special{pa 948 2000}%
\special{pa 934 2020}%
\special{pa 1000 2000}%
\special{fp}%
% STR 2 0 3 0
% 3 600 2300 600 2400 5 0
% $E_{2}$
\put(6.0000,-24.0000){\makebox(0,0){$E_{2}$}}%
% VECTOR 2 0 3 0
% 2 700 2400 1000 2400
% 
\special{pn 8}%
\special{pa 700 2400}%
\special{pa 1000 2400}%
\special{fp}%
\special{sh 1}%
\special{pa 1000 2400}%
\special{pa 934 2380}%
\special{pa 948 2400}%
\special{pa 934 2420}%
\special{pa 1000 2400}%
\special{fp}%
\end{picture}%
&  
\hspace*{1.5cm}
%
%WinTpicVersion3.08
\unitlength 0.1in
\begin{picture}( 18.0000, 17.8500)(  4.0000,-20.0000)
% CIRCLE 1 0 3 0
% 4 800 800 1200 800 1200 800 1200 800
% 
\special{pn 13}%
\special{ar 800 800 400 400  0.0000000 6.2831853}%
% CIRCLE 1 0 3 0
% 4 1800 1600 1400 1600 1400 1600 1400 1600
% 
\special{pn 13}%
\special{ar 1800 1600 400 400  0.0000000 6.2831853}%
% LINE 1 2 3 0
% 2 1100 1100 1470 1380
% 
\special{pn 13}%
\special{pa 1100 1100}%
\special{pa 1470 1380}%
\special{dt 0.045}%
% STR 2 0 3 0
% 3 1090 1000 1090 1100 5 0
% $\bullet$
\put(10.9000,-11.0000){\makebox(0,0){$\bullet$}}%
% STR 2 0 3 0
% 3 1470 1290 1470 1390 5 0
% $\bullet$
\put(14.7000,-13.9000){\makebox(0,0){$\bullet$}}%
% STR 2 0 3 0
% 3 800 200 800 300 5 0
% $C_{1}$
\put(8.0000,-3.0000){\makebox(0,0){$C_{1}$}}%
% STR 2 0 3 0
% 3 1800 1000 1800 1100 5 0
% $C_{2}$
\put(18.0000,-11.0000){\makebox(0,0){$C_{2}$}}%
\end{picture}%
\end{tabular}\\[0.5cm]
\begin{tabular}{ccc}
\makebox[5.25cm][l]{             (a) Subgraph $G_{1}$} &  
\makebox[6.0cm][r]{(b) Subgraph $G_{2}$      } & 
\makebox[5.0cm][r]{(c) Subgraph $G_{3}$           }
\end{tabular}
\end{center}
\caption{(a) a path $P=v_{i_{1}},v_{i_{2}},\ldots,v_{i_{k}}$ that has 
vertices $v_{i_{2}},v_{i_{k-1}} \in S_{1} \cap F_{2}$ 
such that $(v_{i_{2}},v_{i_{3}}) \in E_{1}$ and 
$(v_{i_{k-2}},v_{i_{k-1}}) \in E_{1}$; 
(b) a cycle $C$ and a path $P=v_{i_{1}},v_{i_{2}},\ldots,v_{i_{k}}$ 
incident to $C$ at $v_{i_{k}}$ that has a vertex 
$v_{i_{2}} \in S_{1} \cap F_{2}$ 
such that $(v_{i_{2}},v_{i_{3}}) \in E_{1}$; 
(c) a connected component including cycles $C_{1}$ and $C_{2}$.}
\label{fig-bad-component}
\end{figure}
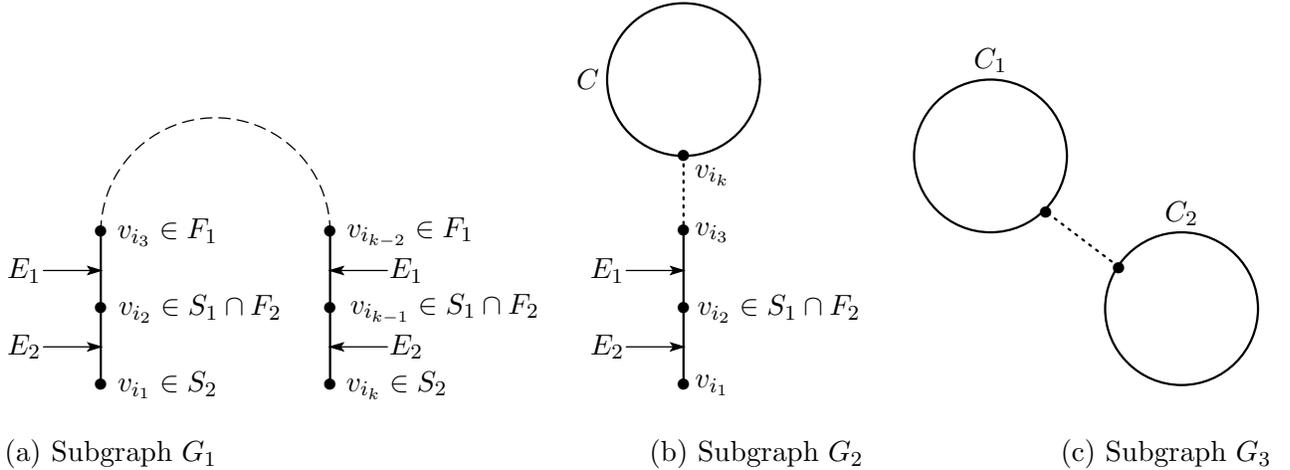
\begin{thm} \label{thm-characterization-two}
An instance $(A,I,\{\vec{\ell}_{x}\}_{x \in A})$ of the strict 
$2$-weighted popular matching problem 
has a well-formed matching iff its fs-relation graph $G=(V,E)$ 
%defined by the procedure in {\rm Figure \ref{fig-graph}}
contains none of the subgraphs $G_{1}$, $G_{2}$, 
nor $G_{3}$ in {\rm Figure \ref{fig-bad-component}}. 
\end{thm}
{\bf Proof:} Assume that 
the graph $G=(V,E)$ contains 
one of the subgraphs $G_{1}$, $G_{2}$, and $G_{3}$ in Figure 
\ref{fig-bad-component}. 
For the case where the graph $G$ contains the subgraph $G_{1}$, 
if the edge $(v_{i_{2}}, v_{i_{3}}) \in E_{1}$ is oriented by 
$v_{i_{2}} \leftarrow v_{i_{3}}$, then the edge 
$(v_{i_{1}}, v_{i_{2}}) \in E_{2}$ is oriented by 
$v_{i_{1}} \leftarrow v_{i_{2}}$ to satisfy the condition (a) of 
Lemma \ref{lemma-orientation}. 
However, this does not meet the condition (c) 
of Lemma \ref{lemma-orientation}, since the vertex 
$v_{i_{2}} \in S_{1} \cap F_{2} \subseteq F_{2}$ has no incoming 
edges in $E_{2}$. 
So the edge $(v_{i_{2}}, v_{i_{3}}) \in E_{1}$ must be 
oriented by $v_{i_{2}} \rightarrow v_{i_{3}}$. It is also the case for  
%OW: ,i.e., ==> ; that is, 
%the edge $(v_{i_{k-2}}, v_{i_{k-1}}) \in E_{1}$, i.e., 
the edge $(v_{i_{k-2}}, v_{i_{k-1}}) \in E_{1}$,
that is,
$(v_{i_{k-2}}, v_{i_{k-1}}) \in E_{1}$ must be oriented by 
$v_{i_{k-1}} \rightarrow v_{i_{k-2}}$. These facts imply that there exists 
$2 < j < k-1$ such that the vertex $v_{i_{j}} \in V$ has 
at least two incoming edges, which violates the condition (a) 
of Lemma \ref{lemma-orientation}. 
%OW: 簡素化
%Thus from Lemma \ref{lemma-orientation}, it 
%follows that if the graph $G$ includes the subgraph $G_{1}$, then an instance 
%$(A,I,\{\vec{\ell}_{x}\}_{x \in A})$ of the strict 2-weighted popular 
%matching problem does not have a well-formed matching. 
%In a way similar to the case where the graph $G$ includes the subgraph $G_{1}$, 
%we can show that if the graph $G$ includes the subgraph $G_{2}$, then 
%an instance $(A,I,\{\vec{\ell}_{x}\}_{x \in A})$ of the strict 
%2-weighted popular matching problem does not have a well-formed matching. 
Thus if the graph $G$ contains the subgraph $G_{1}$,
then the instance does not have a well-formed matching.
Similarly we can show that
if the graph $G$ contains the subgraph $G_{2}$,
then the instance does not have a well-formed matching.
%OW: 簡素化，前の論文の利用
%For the case where the graph $G$ includes the subgraph $G_{3}$,
%we can also show that an instance $(A,I,\{\vec{\ell}_{x}\}_{x \in A})$ of the 
%strict 2-weighted popular matching problem 
%does not have a well-formed matching 
%in a way similar to the argument by Mahdian \cite[Lemma 2]{M.M}.
The case where the graph $G$ contains the subgraph $G_{3}$
can be argued in a way similar to the proof by Mahdian \cite[Lemma 2]{M.M}.
%we can also show that an instance $(A,I,\{\vec{\ell}_{x}\}_{x \in A})$ of the 
%strict 2-weighted popular matching problem 
%does not have a well-formed matching 
%in a way similar to the argument by Mahdian \cite[Lemma 2]{M.M}.

Assume that the graph $G=(V,E)$ does not contain any of the 
subgraph $G_{1}$, $G_{2}$, or $G_{3}$ and let $\{C_{i}\}_{i \geq 1}$ be the 
set of cycles in $G$. We first orient cycles $\{C_{i}\}_{i \geq 1}$. Since 
the graph $G$ does not contain the subgraph $G_{1}$, we can orient 
each cycle $C_{i}$ in one of the clockwise and 
counterclockwise orientations to meet the conditions (a), (b), and (c) 
of Lemma \ref{lemma-orientation}. 
From the assumption that the graph $G$ does not contain the  
subgraph $G_{3}$, the remaining edges can be 
%categorized as follows: Let $E_{\rm tree}^{\rm cyc}$ be the 
%set of edges in subtrees of $G$ 
%that are incident to some cycle 
%$C \in \{C_{i}\}_{i\geq 1}$, and $E_{\rm tree}$ be the set of edges 
%in subtrees of $G$ that are 
%not incident to any cycle $C \in \{C_{i}\}_{i \geq 1}$. 
categorized as follows:
$E_{\rm tree}^{\rm cyc}$ $=$
the set of edges in subtrees of $G$ that are incident
to some cycle $C \in \{C_{i}\}_{i\geq 1}$,
and
$E_{\rm tree}$ $=$ the set of edges in subtrees of $G$
that are not incident to any cycle $C \in \{C_{i}\}_{i \geq 1}$.
Since the graph $G$ does not contain the subgraphs $G_{1}$ 
and $G_{2}$, 
%we can orient the edges in $E_{\rm tree}^{\rm cyc}$ away 
we can orient edges in $E_{\rm tree}^{\rm cyc}$ away 
from the cycles to meet the conditions
(a), (b), and (c) of Lemma \ref{lemma-orientation}.
%We notice that the edges in $E_{\rm tree}$ consist of the set of subtrees 
%$\{T_{j}\}_{j \geq 1}$ of $G$. For each $T \in \{T_{j}\}_{j \geq 1}$, let 
Notice that edges in $E_{\rm tree}$ form subtrees of $G$.
For each such $T$, let $E_{T}^{2}$ be the set of edges $(v,u)$ that is 
assigned to some applicant in $A_{2}$ and $u \in S_{1}\cap F_{2}$. 
For each edge $e=(v,u) \in E_{T}^{2}$, we first orient the 
edge $e$ by $v \rightarrow u$ and then the remaining 
edges in $E_{T}^{2}$ are oriented away 
from each $u \in S_{1} \cap F_{2}$. By the 
assumption that the graph $G$ does not contain the subgraph $G_{1}$, 
such an orientation 
meets the conditions (a), (b), and (c) of Lemma
%\ref{lemma-orientation} for each vertex $v \in T$, and this 
%completes the proof. \BQED\medskip
\ref{lemma-orientation} for each $v \in T$. \BQED\medskip

%This completes the proof. 

From Proposition \ref{prop-mestre} and 
Theorem \ref{thm-characterization-two}, we 
immediately have the following corollary: 
\begin{cor} \label{cor-characterization-two}
Any instance $(A,I,\{\vec{\ell}_{x}\}_{x \in A})$ of the strict $2$-weighted 
popular matching problem with $w_{1}\geq 2w_{2}$ has a $2$-weighted popular matching 
iff its fs-relation graph $G=(V,E)$ 
contains none of the subgraphs $G_{1}$, $G_{2}$, nor $G_{3}$ 
in {\rm Figure \ref{fig-bad-component}}. 
\end{cor}

% omit

%OW: ここから新しい版
Let us consider a random instance of
the complete and strict 2-weighted popular matching problem. 
Roughly speaking, a natural uniform distribution is considered here.
That is, given a set $A=A_1\cup A_2$ of $n$ applicants and a set $I$ of $m$ items, 
and we consider an instance obtained by defining a random preference 
list $\vec{\ell}_{x}$ for each applicant $x \in A$, 
which is a permutation on $I$ that is chosen independently and 
uniformly at random. But as discussed above 
for the 2-weighted case, the situation is completely determined
by the corresponding fs-relation graph that depends only on 
the first and second items of applicants.
Thus, instead of considering a random instance of the problem,
we simply define the first and second items as follows,
and discuss with
the fs-relation graph $G=(V,E)$ obtained defined by $f_{1}$-, 
$s_1$-, $f_2$-, and $s_2$-items.
\begin{namelist}{   (4)}
\item[(1)]
For each $x\in A_1$,
assign an item $p\in I$ as a $f_1$-item $f_1(x)$ 
independently and uniformly at random, 
and let $F_1$ be the set of all $f_1$-items;
\item[(2)]
For each $x\in A_1$,
assign an item $p\in I-F_1$ as a $s_1$-item $s_1(x)$ 
independently and uniformly at random, 
and let $S_1$ be the set of all $s_1$-items;
\item[(3)]
For each $x\in A_2$,
assign an item $p\in I-F_1$ as a $f_2$-item $f_2(x)$ 
independently and uniformly at random, 
and let $F_2$ be the set of all $f_2$-items; and
\item[(4)]
For each $x\in A_2$,
assign an item $p\in I-(F_1\cup F_2)$ as a $s_2$-item $s_2(x)$ 
independently and uniformly at random, 
and let $S_2$ be the set of all $s_2$-items.
\end{namelist}
It is easy to see that
this choice of first and second items is
the same as defining first and second items~from a random instance of the 
complete and strict 2-weighted popular matching problem.
\OMIT{
From Corollary \ref{cor-characterization-two}, it follows that 
a random choice of an instance 
$(A,I,\{\vec{\ell}_{x}\}_{x \in A})$ of the complete and strict 
2-weighted popular matching problem is equivalent to a 
random choice of an instance $G=(V,E)$ of the fs-relation graphs 
(by the process in Figure \ref{fig-random-graph}). 
In the rest of this paper, we consider a random 
instance $G=(V,E)$ of the fs-relation graphs 
(by the process in Figure \ref{fig-random-graph}) 
instead of a random instance 
$(A,I,\{\vec{\ell}_{x}\}_{x \in A})$ 
of the complete and strict 2-weighted popular matching problem.  }
%
%
%==================================================================
\section{Lower Bounds for the 2-Weighted Popular Matching Problem}
\label{sec-lower}
\begin{figure}[ht]
\begin{center}
\hspace*{-3.7cm}
%
%WinTpicVersion3.08
\unitlength 0.1in
\begin{picture}( 36.3000, 15.8500)( -3.3000,-18.0000)
% LINE 1 0 3 0
% 2 2200 800 2800 1200
% 
\special{pn 13}%
\special{pa 2200 800}%
\special{pa 2800 1200}%
\special{fp}%
% LINE 1 0 3 0
% 2 2200 800 1600 1200
% 
\special{pn 13}%
\special{pa 2200 800}%
\special{pa 1600 1200}%
\special{fp}%
% LINE 1 0 3 0
% 2 1600 1200 1600 1800
% 
\special{pn 13}%
\special{pa 1600 1200}%
\special{pa 1600 1800}%
\special{fp}%
% LINE 1 0 3 0
% 2 2800 1200 2800 1800
% 
\special{pn 13}%
\special{pa 2800 1200}%
\special{pa 2800 1800}%
\special{fp}%
% STR 2 0 3 0
% 3 2200 700 2200 800 5 0
% $\bullet$
\put(22.0000,-8.0000){\makebox(0,0){$\bullet$}}%
% STR 2 0 3 0
% 3 1600 1100 1600 1200 5 0
% $\bullet$
\put(16.0000,-12.0000){\makebox(0,0){$\bullet$}}%
% STR 2 0 3 0
% 3 2800 1100 2800 1200 5 0
% $\bullet$
\put(28.0000,-12.0000){\makebox(0,0){$\bullet$}}%
% STR 2 0 3 0
% 3 800 1400 800 1500 5 0
% $y_{1} \in A_{2}$
\put(8.0000,-15.0000){\makebox(0,0){$y_{1} \in A_{2}$}}%
% STR 2 0 3 0
% 3 2800 500 2800 600 5 0
% $x_{2} \in E_{1}$
\put(28.0000,-6.0000){\makebox(0,0){$x_{2} \in E_{1}$}}%
% STR 2 0 3 0
% 3 3600 1400 3600 1500 5 0
% $y_{2} \in A_{2}$
\put(36.0000,-15.0000){\makebox(0,0){$y_{2} \in A_{2}$}}%
% STR 2 0 3 0
% 3 1600 500 1600 600 5 0
% $x_{1} \in A_{1}$
\put(16.0000,-6.0000){\makebox(0,0){$x_{1} \in A_{1}$}}%
% VECTOR 2 0 3 0
% 2 1700 700 1900 1000
% 
\special{pn 8}%
\special{pa 1700 700}%
\special{pa 1900 1000}%
\special{fp}%
\special{sh 1}%
\special{pa 1900 1000}%
\special{pa 1880 934}%
\special{pa 1870 956}%
\special{pa 1846 956}%
\special{pa 1900 1000}%
\special{fp}%
% VECTOR 2 0 3 0
% 2 2700 700 2500 1000
% 
\special{pn 8}%
\special{pa 2700 700}%
\special{pa 2500 1000}%
\special{fp}%
\special{sh 1}%
\special{pa 2500 1000}%
\special{pa 2554 956}%
\special{pa 2530 956}%
\special{pa 2520 934}%
\special{pa 2500 1000}%
\special{fp}%
% VECTOR 2 0 3 0
% 2 1100 1500 1600 1500
% 
\special{pn 8}%
\special{pa 1100 1500}%
\special{pa 1600 1500}%
\special{fp}%
\special{sh 1}%
\special{pa 1600 1500}%
\special{pa 1534 1480}%
\special{pa 1548 1500}%
\special{pa 1534 1520}%
\special{pa 1600 1500}%
\special{fp}%
% VECTOR 2 0 3 0
% 2 3300 1500 2800 1500
% 
\special{pn 8}%
\special{pa 3300 1500}%
\special{pa 2800 1500}%
\special{fp}%
\special{sh 1}%
\special{pa 2800 1500}%
\special{pa 2868 1520}%
\special{pa 2854 1500}%
\special{pa 2868 1480}%
\special{pa 2800 1500}%
\special{fp}%
% STR 2 0 3 0
% 3 750 950 750 1050 5 0
% $q \in S_{1} \cap F_{2}$
\put(7.5000,-10.5000){\makebox(0,0){$q \in S_{1} \cap F_{2}$}}%
% STR 2 0 3 0
% 3 3650 950 3650 1050 5 0
% $r \in S_{1} \cap F_{2}$
\put(36.5000,-10.5000){\makebox(0,0){$r \in S_{1} \cap F_{2}$}}%
% VECTOR 2 0 3 0
% 2 1200 1100 1600 1200
% 
\special{pn 8}%
\special{pa 1200 1100}%
\special{pa 1600 1200}%
\special{fp}%
\special{sh 1}%
\special{pa 1600 1200}%
\special{pa 1540 1164}%
\special{pa 1548 1188}%
\special{pa 1530 1204}%
\special{pa 1600 1200}%
\special{fp}%
% VECTOR 2 0 3 0
% 2 3200 1100 2800 1200
% 
\special{pn 8}%
\special{pa 3200 1100}%
\special{pa 2800 1200}%
\special{fp}%
\special{sh 1}%
\special{pa 2800 1200}%
\special{pa 2870 1204}%
\special{pa 2852 1188}%
\special{pa 2860 1164}%
\special{pa 2800 1200}%
\special{fp}%
% STR 2 0 3 0
% 3 2200 200 2200 300 5 0
% $p \in F_{1}$
\put(22.0000,-3.0000){\makebox(0,0){$p \in F_{1}$}}%
% VECTOR 2 0 3 0
% 2 2200 400 2200 800
% 
\special{pn 8}%
\special{pa 2200 400}%
\special{pa 2200 800}%
\special{fp}%
\special{sh 1}%
\special{pa 2200 800}%
\special{pa 2220 734}%
\special{pa 2200 748}%
\special{pa 2180 734}%
\special{pa 2200 800}%
\special{fp}%
\end{picture}%
\end{center}
\caption{The Simplest ``Bad'' Subgraphs $G_{1}'$} 
\label{fig-small-subgraph}
\end{figure}
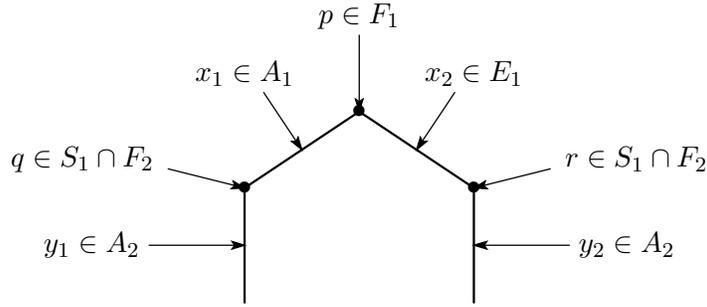

%==================================================================
%
Let $n$ be the number of applicants and $m$ be the number of items.
%Let $m = \beta n$, where $\beta$ could be a function of $n$, and 
Assume that $m$ is large enough so that 
$m-n\geq m/c$ for some constant $c>1$, i.e., $m \geq cn/(c-1)$. 
For any constant $0 < \delta < 1$, let $n_{1}=\delta n$ and 
$n_{2}=(1-\delta) n$ be the numbers of applicants in $A_{1}$ and $A_{2}$, 
respectively. 
% be the number of applicants in $A_{2}$. 
In this section, we show a lower bound for $m$ such that a random instance of 
the complete and strict 2-weighted popular matching problem 
has a $2$-weighted popular matching with low probability. 
\begin{thm} \label{thm-lower}
%
%Let $m = \beta n$. 
If $m/n^{4/3}=o(1)$, then a 
random instance of the complete and strict 
$2$-weighted popular matching problem with $w_{1}\geq 2w_{2}$ has 
a $2$-weighted popular matching with probability $o(1)$. 
\end{thm}
%
%OW: 上のランダムインスタンスの説明の変更にともなう変更
%
%OW: 旧版 omitted
\OMIT{
{\bf Proof:} Let $F_{1},F_{2}$ be the set of the first items for 
applicants in $A_{1},A_{2}$, respectively, and $S_{1},S_{2}$ be the 
set of the second items for 
applicants in $A_{1},A_{2}$, respectively. By the definitions of 
$F_{1},F_{2},S_{1},S_{2}$, we have that 
$F_{1} \cap S_{1}=\emptyset$; 
$F_{1} \cap F_{2}=\emptyset$; 
$F_{1} \cap S_{2}=\emptyset$; 
$F_{2} \cap S_{2}=\emptyset$, but we may have that 
$S_{1} \cap F_{2} \neq \emptyset$ or $S_{1} \cap S_{2} \neq \emptyset$. 
Let $R_{1}=I-F_{1}$ and $R_{2}=R_{1}-F_{2}=I-(F_{1}\cup F_{2})$. 
It is obvious that 
$1 \leq \leng{F_{1}} \leq \delta n$ and  
$1 \leq \leng{F_{2}} \leq (1-\delta) n$, which implies that 
$m- \delta n \leq \leng{R_{1}} \leq m$ and  
$m- n \leq \leng{R_{2}} \leq m$. It follows from Corollary 
\ref{cor-characterization-two} that 
an instance $(A,I,\{\vec{\ell}_{x}\}_{x \in A})$ of the 
complete and strict 2-weighted popular matching problem with $w_{1}\geq 2w_{2}$ 
does not have a 2-weighted popular 
matching iff its fs-relation graph $G=(V,E)$ includes one 
of the bad subgraphs $G_{1}$, $G_{2}$, and $G_{3}$ in Figure \ref{fig-bad-component}.
To show the theorem, it suffices to consider the probability that  
a random instance $G=(V,E)$ of the fs-relation graphs includes 
the simplest bad
subgraphs $G_{1}'$ as shown in Figure \ref{fig-small-subgraph}. } % omit
%
%OW: ここから新版
{\bf Proof:}
Consider a random fs-relation graph $G=(V,E)$.
As shown in Corollary \ref{cor-characterization-two},
it suffices to prove that $G=(V,E)$ contains 
one of the graphs $G_{1}$, $G_{2}$, and $G_{3}$ of Figure \ref{fig-bad-component}
with high probability.
But~here~we focus on
one simple such graph, namely, $G_{1}'$ given Figure \ref{fig-small-subgraph},
and in the following,
we~argue~that~the~probabil\-ity that $G=(V,E)$ contains $G_1'$ is high
if $m/n^{4/3}=o(1)$. 

Let $F_{1}$ and $F_{2}$ be
the sets of the first items,
$S_{1}$ and $S_{2}$ be the sets of the second items, respectively,
for applicants in $A_{1}$ and $A_{2}$.
By the definitions of 
$F_{1}$, $F_{2}$, $S_{1}$, and $S_{2}$,
we have that $F_{1} \cap S_{1}=\emptyset$, 
$F_{1} \cap F_{2}=\emptyset$, 
$F_{1} \cap S_{2}=\emptyset$, and $F_{2} \cap S_{2}=\emptyset$.
On the other hand,
we may have that 
$S_{1} \cap F_{2} \neq \emptyset$ or $S_{1} \cap S_{2} \neq \emptyset$.
Let $R_{1}=I-F_{1}$ and $R_{2}=R_{1}-F_{2}=I-(F_{1}\cup F_{2})$.
It is obvious that
$1 \leq \leng{F_{1}} \leq \delta n$ and 
$1 \leq \leng{F_{2}} \leq (1-\delta) n$,
which implies that $m- \delta n \leq \leng{R_{1}} \leq m$ 
and $m- n \leq \leng{R_{2}} \leq m$. 

For any pair of $x_{1},x_{2} \in A_{1}$ such that $x_{1}<x_{2}$ and any 
pair of $y_{1},y_{2} \in A_{2}$ such that $y_{1}\neq y_{2}$,~we~simply use 
$\vec{v}$ to denote $(x_{1},x_{2},y_{1},y_{2})$, and  
$T$ to denote the set of all such $\vec{v}$'s. 
Since $n_{1}=\delta n=\leng{A_{1}}$~and~$n_{2}=(1-\delta) n=\leng{A_{2}}$, 
we have that for sufficiently large $n$, 
\begin{equation}
\leng{T} = \cmb{n_{1}}{2} n_{2}(n_{2}-1) \geq
\frac{\delta^{2}(1-\delta)^{2}}{3}n^{4}. \label{eq-N-lower}
%
%\leng{T} & = & \cmb{n_{1}}{2} n_{2}(n_{2}-1) \leq
%\frac{\delta^{2}(1-\delta)^{2}}{2}n^{4}. \label{eq-N-upper}
%
\end{equation}
For each $\vec{v} =(x_{1},x_{2},y_{1},y_{2}) \in T$, 
%any pair of edges $x_{1},x_{2} \in A_{1}$ 
%such that $x_{1}<x_{2}$ and any pair of edges $y_{1},y_{2} \in A_{2}$, 
define a random variable $Z_{\vec{v}}$ 
to be $Z_{\vec{v}}=1$ if 
$x_{1},x_{2},y_{1}$, and $y_{2}$ form 
the bad subgraph $G_{1}'$ in Figure \ref{fig-small-subgraph}; 
$Z_{\vec{v}}=0$ otherwise. 
Let $Z=\sum_{\vec{v} \in V} Z_{\vec{v}}$. 
Then from Chebyshev's Inequality \cite[Theorem 3.3]{MR},  
it follows that 
%, i.e., $\Pr[\leng{Z-{\bf E}[Z]} \geq t \sigma_{Z}]\leq 1/t^{2}$, 
%it follows that 
%
\begin{eqnarray}
\Pr \left[ Z = 0 \right] & \leq & 
\Pr\left[\leng{Z - {\bf E}[Z]} \geq {\bf E}[Z]\right]\nonumber\\
& = & \Pr\left[\leng{Z - {\bf E}[Z]} \geq 
\frac{{\bf E}[Z]}{\sigma_{Z}} \sigma_{Z} \right] 
\leq \frac{\sigma_{Z}^{2}}{{\bf E}^{2}[Z]} = 
\frac{{\bf Var}[Z]}{{\bf E}^{2}[Z]}. \label{eq-chebyshev}
\end{eqnarray}
To derive the lower bound for $\Pr[Z>0]$, we estimate the upper 
bound for ${\bf Var}[Z]/{\bf E}^{2}[Z]$. 
We~first~consider ${\bf E}[Z]$. For each $\vec{v} \in T$, 
it is easy to see that 
\begin{eqnarray}
\Pr\left[Z_{\vec{v}}=1\right] & \geq & \frac{1}{m} \cdot 
\left(\frac{1}{m}\right)^{2} = \frac{1}{m^{3}};\nonumber\\ 
\Pr\left[Z_{\vec{v}}=1\right] & \leq & \frac{1}{m} \cdot 
\left(\frac{1}{m-n_{1}}\right)^{2}
\leq \frac{1}{m} \cdot \left(\frac{1}{m-n}\right)^{2}
= \frac{c^{2}}{m^{3}}, \label{eq-Zv-upper}
\end{eqnarray}
where Inequality (\ref{eq-Zv-upper}) follows from the assumption that 
$m-n_{1} \geq m-n \geq m/c$ for some constant $c>1$. Thus from the 
estimations for $\Pr[Z_{\vec{v}}=1]$, it follows that 
\begin{eqnarray}
{\bf E}\left[Z\right] & = &
{\bf E}\left[\sum_{\vec{v} \in T}Z_{\vec{v}}\right]= 
\sum_{\vec{v} \in T} {\bf E}\left[Z_{\vec{v}}\right] =
\sum_{\vec{v} \in T} \Pr \left[Z_{\vec{v}}=1\right] 
\geq \frac{\leng{T}}{m^{3}}; \label{eq-Z-lower}\\
{\bf E}\left[Z\right] & = &
{\bf E}\left[\sum_{\vec{v} \in T}Z_{\vec{v}}\right]= 
\sum_{\vec{v} \in T} {\bf E}\left[Z_{\vec{v}}\right]=
\sum_{\vec{v} \in T} \Pr\left[Z_{\vec{v}}=1\right] 
\leq \frac{c^{2}\leng{T}}{m^{3}}. \label{eq-Z-upper}
\end{eqnarray}
%
%where Inequality (\ref{eq-Z-lower}) follows from 
%Inequality (\ref{eq-Zv-lower}) and 
%Inequality (\ref{eq-Z-upper}) follows from 
%Inequality (\ref{eq-Zv-upper}). 
We then consider ${\bf Var}[Z]$. From the definition of 
${\bf Var}[Z]$, it follows that 
\begin{eqnarray}
{\bf Var}[Z] & = &
{\bf E}\left[\left(\sum_{\vec{v}\in T} Z_{\vec{v}}\right)^{2}\right] 
- \left({\bf E}\left[\sum_{\vec{v}\in T} Z_{\vec{v}}\right]\right)^{2}\nonumber\\
& = & {\bf E}\left[\sum_{\vec{v}\in T} Z_{\vec{v}}^{2} + 
\sum_{\vec{v} \in T} \sum_{\vec{w} \in T-\{\vec{v}\}} Z_{\vec{v}} Z_{\vec{w}}
\right]- 
\left({\bf E}\left[\sum_{\vec{v}\in T} Z_{\vec{v}}\right]\right)^{2}\nonumber\\
& = & {\bf E}\left[\sum_{\vec{v}\in T} Z_{\vec{v}}\right]  
- \left({\bf E}\left[\sum_{\vec{v}\in T} Z_{\vec{v}}\right]\right)^{2}
+ \sum_{\vec{v} \in T} \sum_{\vec{w} \in T-\{\vec{v}\}} {\bf E}
\left[ Z_{\vec{v}} Z_{\vec{w}}\right]\nonumber\\
& = & {\bf E}[Z] - {\bf E}^{2}[Z] + 
\sum_{\vec{v} \in T} \sum_{\vec{w} \in T-\{\vec{v}\}} {\bf E}
\left[ Z_{\vec{v}} Z_{\vec{w}}\right]. \label{eq-Var-Z-upper}
%& = & 
%\frac{c^{2}N}{m^{3}} - 
%\left(\frac{N}{m^{3}}\right)^{2} 
%+ \sum_{\vec{v} \in T} \sum_{\vec{w} \in T-\{\vec{v}\}} {\bf E}
%\left[ Z_{\vec{v}} Z_{\vec{w}}\right], 
%
\end{eqnarray}
%
%where Inequality (\ref{eq-Var-Z-upper}) follows from Inequalities 
%(\ref{eq-Z-lower}) and (\ref{eq-Z-upper}). 
%
In the following, we estimate the last term 
of Equality (\ref{eq-Var-Z-upper}). 
For each $\vec{v}=(x_{1},x_{2},y_{1},y_{2}) \in T$ and each $0 \leq h \leq 2$,
we say that 
$\vec{w}=(x_{1}',x_{2}',y_{1}',y_{2}') \in T - \{\vec{v}\}$ is 
$h$-common to $\vec{v}$ if $\leng{\{x_{1},x_{2}\} \cap 
\{x_{1}',x_{2}'\}}=h$. For any $\vec{w}=(x_{1}',x_{2}',y_{1}',y_{2}') \in T$ 
that is 2-common to 
$\vec{v}$, we have that $x_{1}=x_{1}'$ and $x_{2}=x_{2}'$, 
because if $x_{1}=x_{2}'$ and $x_{2}=x_{1}'$, then $x_{1}=x_{2}'>x_{1}'=x_{2}$, 
which contradicts the assumption that $x_{1}<x_{2}$. For each $\vec{v}\in T$, 
we use $T_{2}(\vec{v})$ to denote the set of 
$\vec{w} \in T-\{\vec{v}\}$ that is 2-common to $\vec{v}$; 
$T_{1}(\vec{v})$ to denote the set of 
$\vec{w} \in T-\{\vec{v}\}$ that is 1-common to $\vec{v}$; 
$T_{0}(\vec{v})$ to denote the set of 
$\vec{w} \in T-\{\vec{v}\}$ that is 
0-common~to~$\vec{v}$.~Then~from~the~assumption that $m-n\geq m/c$, 
it follows that 
\begin{eqnarray}
\sum_{\vec{v} \in T} 
\sum_{\vec{w} \in T_{2}(\vec{v})} 
{\bf E} \left[ Z_{\vec{v}} Z_{\vec{w}}\right] 
& \leq & 
\left\{\frac{c^{4}(1-\delta)^{2}}{m^{5}} n^{2}
+ 
\frac{2c^{3}(1-\delta)}{m^{4}} n \right\} \leng{T}; \label{eq-2-common}\\
\sum_{\vec{v} \in T} 
\sum_{\vec{w} \in T_{1}(\vec{v})} 
{\bf E} \left[ Z_{\vec{v}} Z_{\vec{w}}\right] 
& \leq & 
\left\{
\frac{4c^{4}\delta(1-\delta)^{2}}{m^{6}}n^{3}+ 
\frac{4c^{3}\delta(1-\delta)}{m^{5}}n^{2}+ 
\frac{4c^{3}\delta}{m^{5}}n \right\} \leng{T};\label{eq-1-common}\\
\sum_{\vec{v} \in T} 
\sum_{\vec{w} \in T_{0}(\vec{v})} {\bf E} \left[ Z_{\vec{v}} Z_{\vec{w}}\right] 
& \leq & {\bf E}^{2}[Z] 
+ 
\left\{ \frac{2c^{4}\delta^{2}(1-\delta)}{m^{6}}n^{3} + 
\frac{c^{4}\delta^{2}}{m^{6}}n^{2} \right\}\leng{T}. \label{eq-0-common}
\end{eqnarray}
The proofs of Inequalities (\ref{eq-2-common}), 
(\ref{eq-1-common}), and (\ref{eq-0-common}) are shown in Subsections 
\ref{derivations-inequality-2-common}, 
\ref{derivations-inequality-1-common}, and 
\ref{derivations-inequality-0-common}, respectively. 
Thus from Inequalities (\ref{eq-Z-upper}), 
(\ref{eq-Var-Z-upper}), (\ref{eq-2-common}), 
(\ref{eq-1-common}), and (\ref{eq-0-common}), it follows that 
\begin{eqnarray}
\lefteqn{{\bf Var}[Z] \leq {\bf E}[Z]-{\bf E}^{2}[Z]+ 
\sum_{\vec{v} \in T} \sum_{\vec{w} \in T-\{\vec{v}\}} {\bf E}
\left[ Z_{\vec{v}} Z_{\vec{w}}\right]}\nonumber\\
& =  & {\bf E}[Z]-{\bf E}^{2}[Z] 
+ \sum_{\vec{v} \in T} \sum_{\vec{w} \in T_{2}(\vec{v})} {\bf E}
\left[ Z_{\vec{v}} Z_{\vec{w}}\right] 
+ \sum_{\vec{v} \in T} \sum_{\vec{w} \in T_{1}(\vec{v})} {\bf E}
\left[ Z_{\vec{v}} Z_{\vec{w}}\right] 
+ \sum_{\vec{v} \in T} \sum_{\vec{w} \in T_{0}(\vec{v})} {\bf E}
\left[ Z_{\vec{v}} Z_{\vec{w}}\right]\nonumber\\
& \leq & \frac{c^{2}\leng{T}}{m^{3}}
+ \left\{\frac{c^{4}(1-\delta)^{2}}{m^{5}} n^{2}
+ 
\frac{2c^{3}(1-\delta)}{m^{4}} n \right\} \leng{T}\nonumber\\
& &                + 
\left\{
\frac{4c^{4}\delta(1-\delta)^{2}}{m^{6}}n^{3}+ 
\frac{4c^{3}\delta(1-\delta)}{m^{5}}n^{2}+ 
\frac{4c^{3}\delta}{m^{5}}n \right\} \leng{T}\nonumber\\
& &                +  
\left\{ \frac{2c^{4}\delta^{2}(1-\delta)}{m^{6}}n^{3} + 
\frac{c^{4}\delta^{2}}{m^{6}}n^{2} \right\}\leng{T}\nonumber\\
& \leq & \frac{c^{2}\leng{T}}{m^{3}}
\Biggl\{1+ \frac{c^{2}(1-\delta)^{2}}{m^{2}} n^{2}
+ 
\frac{2c(1-\delta)}{m} n 
+ 
\frac{4c^{2}\delta(1-\delta)^{2}}{m^{3}}n^{3}\nonumber\\
& &                +  
\frac{4c\delta(1-\delta)}{m^{2}}n^{2}+ 
\frac{4c \delta}{m^{2}} n 
+ \frac{2c^{2}\delta^{2}(1-\delta)}{m^{3}}n^{3} + 
\frac{c^{2}\delta^{2}}{m^{3}}n^{2} \Biggr\}\nonumber\\
& \leq & \frac{c^{2}\leng{T}}{m^{3}} 
\Biggl\{1+ (c-1)^{2}(1-\delta)^{2}+ 
2(c-1)(1-\delta)+ 
\frac{4(c-1)^{3}\delta(1-\delta)^{2}}{c}\nonumber\\
& &                 + 
\frac{4(c-1)^{2}\delta(1-\delta)}{c}+ 
\frac{4(c-1)\delta}{m}+ 
\frac{2(c-1)^{3}\delta^{2}(1-\delta)}{c} + 
\frac{(c-1)^{2}\delta^{2}}{m} \Biggr\} \label{eq-Var-Z-upper-1},
\end{eqnarray}
where Inequality (\ref{eq-Var-Z-upper-1}) follows from the assumption 
that $m-n\geq m/c$, i.e., $cn/m \leq c-1$.~Thus~it~follows~that 
${\bf Var}[Z] \leq d \leng{T}/m^{3}$ for some constant $d$ that is 
determined by the constants $0 < \delta < 1$~and~$c>1$.~Then~from 
Inequalities (\ref{eq-N-lower}), (\ref{eq-chebyshev}),  and
(\ref{eq-Z-lower}), %and (\ref{eq-Var-Z-upper-1}), 
we finally have that 
\[
\Pr\left[Z=0\right] \leq \frac{{\bf Var}[Z]}{{\bf E}^{2}[Z]} 
\leq \frac{d\leng{T}}{m^{3}}\cdot \frac{m^{6}}{\leng{T}^{2}} = 
\frac{dm^{3}}{\leng{T}}
\leq \frac{3dm^{3}}{\delta^{2}(1-\delta)^{2}n^{4}}
= O\left(\frac{m^{3}}{n^{4}}\right), 
%= \frac{2d}{\delta^{2}(1-\delta)^{2}}\cdot \frac{m^{3}}{n^{4}},
%
\]
which implies that $\Pr[Z=0]=o(1)$ for any $m \geq n$ with 
$m/n^{4/3}=o(1)$.
Therefore, if $m/n^{4/3}=o(1)$,~then with probability $1-o(1)$, 
we have $Z>0$, that is,
$G=(V,E)$ contains $G'_1$ as a subgraph. \BQED
%
%==================================================================
\section{Upper Bounds for the 2-Weighted Popular Matching Problem}
\label{sec-upper}
%==================================================================
%
As shown in Theorem \ref{thm-lower}, 
a random instance of the complete and strict 2-weighted popular matching~prob\-lem  
has a 2-weighted popular matching with probability $o(1)$ 
if $m/n^{4/3}=o(1)$. Here we consider roughly opposite case, i.e., $n^{4/3}/m=o(1)$,
and prove that a random instance has a 2-weighted popular matching 
with probability $1-o(1)$. 
%In this section, we derive upper bounds for 
%$m$ such that a random instance of the complete and strict 
%2-weighted popular matching problem 
%has a 2-weighted popular matching with high probability.

First we show
the following lemma that will greatly simplify our later analysis.
%The following lemma plays a crucial role to 
%derive upper bounds for $m$. 
%
\begin{lemma} \label{lemma-no-cycle}
If $n/m=o(1)$,
then a random instance $G=(V,E)$ of the fs-relation graphs
contains~a~cycle as a subgraph with probability $o(1)$.
\end{lemma}
{\bf Proof:} For each $\ell \geq 2$, let $C_{\ell}$ be 
a cycle with $\ell$ vertices and $\ell$ edges, and
%OW: Figure 2 無しで済ます
${\cal E}_{\ell}^{\rm cyc}$ be the event that a random fs-relation graph $G=(V,E)$
%${\cal E}_{\ell}^{\rm cyc}$ be the event that a random instance $G=(V,E)$ of 
%the fs-relation graphs generated by the process in Figure \ref{fig-random-graph} 
%for the complete and strict 2-weighted popular matching problem 
contains a cycle $C_{\ell}$. 
Then from the assumption that $m-n\geq m/c$ for some constant $c >1$, 
it follows that 
\begin{eqnarray*}
\lefteqn{\Pr\left[G \mbox{ contains a cycle}\right] = 
\Pr\left[\bigcup_{\ell \geq 2} {\cal E}_{\ell}^{\rm cyc}\right] \leq 
\sum_{\ell \geq 2} \Pr [{\cal E}_{\ell}^{\rm cyc}]}\\
& \leq & \sum_{\ell \geq 2} \left\{
\frac{1}{2\ell} \ell! \cmb{m}{\ell}  
\ell! \cmb{n}{\ell} \left(\frac{1}{m-n}\right)^{2\ell} \right\}
\leq \sum_{\ell \geq 2} \left\{
\frac{1}{2\ell} m^{\ell} n^{\ell}  
\left(\frac{c}{m}\right)^{2\ell}\right\}\\
& = & \sum_{\ell \geq 2} 
\frac{1}{2\ell} \left(\frac{c^{2}n}{m}\right)^{\ell} 
\leq \sum_{\ell \geq 2} 
\left(\frac{c^{2}n}{m}\right)^{\ell}= 
\frac{c^{4}n^{2}}{m^{2}} \sum_{h\geq 0} \left(\frac{c^{2}n}{m}\right)^{h} = 
O\left(\frac{n^{2}}{m^{2}}\right), 
\end{eqnarray*}
where the last equality follows from the assumption that 
$n/m=o(1)$ and $c>1$ is a constant.~Thus~it~follows that if $n/m=o(1)$, then 
%OW: Figure 2 無しで済ます
a random fs-relation graph $G=(V,E)$
%a random instance $G=(V,E)$ of the fs-relation graphs 
%generated by the process in Figure \ref{fig-random-graph}
%for the complete and strict 2-weighted popular matching problem 
contains a cycle as a subgraph with probability $o(1)$. \BQED
\begin{thm} \label{thm-upper}
If $n^{4/3}/m = o(1)$, then a 
random instance of the complete and strict $2$-weighted popular matching problem 
with $w_{1}\geq 2w_{2}$ has 
a $2$-weighted popular matching with probability $1 - o(1)$. 
\end{thm}
{\bf Proof:}
Consider a random fs-relation graph $G=(V,E)$ corresponding to 
a random instance~of~the~complete and strict $2$-weighted popular matching problem. 
By Lemma \ref{lemma-no-cycle} and the assumption~that~$n^{4/3}/m=o(1)$, 
we know that the fs-relation graph 
$G=(V,E)$ contains bad subgraphs $G_{2}$ or $G_{3}$ of 
Figure~\ref{fig-bad-component}~with~vanishing probability $o(1)$. 
Thus in the rest of the proof, we estimate the probability that the 
graph $G=(V,E)$ contains a bad subgraph
$G_{1}$ of Figure \ref{fig-bad-component}. 

For any $\ell \geq 4$, let $P_{\ell}$ be a path 
with $\ell+1$ vertices 
%OW: 簡素化＆ Figure 2 無しで済ませる
%and $\ell$ edges, and ${\cal E}_{\ell}^{\rm path}$ be the event that a 
%random instance $G=(V,E)$ of the fs-relation graphs 
%generated by the process in Figure \ref{fig-random-graph} 
%for the complete and strict 2-weighted popular matching problem 
and $\ell$ edges, and ${\cal E}_{\ell}^{\rm path}$ be the event that $G=(V,E)$
contains a path 
$P_{\ell}$. It is obvious that a path $P_{\ell}$ 
is a bad subgraph $G_{1}$ for each $\ell \geq 4$. 
Then~from~the~as\-sumption that $m-n\geq m/c$ for some constant $c>1$, 
it follows that 
\begin{eqnarray*}
\lefteqn{\Pr\left[G \mbox{ contains a bad subgraph $G_{1}$}\right]
= \Pr \left[\bigcup_{\ell \geq 4} {\cal E}_{\ell}^{\rm path}\right]
\leq \sum_{\ell\geq 4} \Pr\left[ {\cal E}_{\ell}^{\rm path}\right]}\\
& \leq & \sum_{\ell \geq 4} \left\{\frac{1}{(m-n)^{2\ell}} 
(\ell+1)! \cmb{m}{\ell+1} \ell! \cmb{n}{\ell}\right\}\\
& \leq & \sum_{\ell\geq 4} \left\{\left(\frac{c}{m}\right)^{2\ell} 
m^{\ell+1} n^{\ell} \right\} = 
\frac{c^{8}n^{4}}{m^{3}} \sum_{h\geq 0} \left(\frac{c^{2}n}{m} \right)^{h} = 
O\left(\frac{n^{4}}{m^{3}}\right), 
\end{eqnarray*}
where the last equality follows from the assumption that $n^{4/3}/m=o(1)$ 
and that $c>1$ is a constant.~Notice that $n/m=o(1)$ if $n^{4/3}/m=o(1)$. 
Thus from Lemma \ref{lemma-no-cycle} and 
Corollary \ref{cor-characterization-two}, 
it follows that if $n^{4/3}/m = o(1)$, then a random instance of 
the complete and strict 2-weighted popular matching problem~has~a~2-weighted popular matching 
with probability $1-o(1)$. \BQED
%
%===========================================
\section{Concluding Remarks} \label{remarks}
%===========================================
%
In this paper, we have analyzed the 
%complete and strict 
2-weighted matching problem, and have shown that 
(Theorem \ref{thm-lower}) if $m/n^{4/3}=o(1)$, 
then a random instance of the complete and strict 2-weighted 
popular matching problem with $w_{1}\geq 2w_{2}$ has a 
2-weighted popular matching with probability $o(1)$; (Theorem \ref{thm-upper}) 
if $n^{4/3}/m=o(1)$, 
then a random instance of the complete and strict 2-weighted 
popular matching problem with $w_{1}\geq 2w_{2}$ has a 
2-weighted popular matching with probability $1-o(1)$. These results imply 
that~there~exists~a~threshold $m \approx n^{4/3}$ 
to admit 2-weighted popular matchings, % with probability $1-o(1)$, 
which is quite different from the case for the %complete and strict 
single category popular matching problem due to
Mahdian \cite{M.M}.
%OW: 少し書き直してみました．

%OW: 旧版
\OMIT{
We can generalize Theorem \ref{thm-lower} to any integer $k \geq 2$, i.e., 
%
%In a way similar to the proof of Theorem \ref{thm-lower}, 
%we can show the following lower bounds for $m$ such that 
%for any integer $k\geq 2$, 
%a random instance of the complete and strict
%$k$-weighted matching problem does not have a $k$-weighted popular matching with 
%high probability, i.e.,  
%
\begin{thm} \label{thm-lower-k}
For any integer $k > 2$, if $m/n^{4/3}=o(1)$, then a 
random instance of the complete and strict $k$-weighted popular 
matching problem with $w_{i} \geq 2w_{i+1}$ $(1 \leq i \leq k-1)$ 
has a $k$-weighted popular matching with probability $o(1)$. 
\end{thm}
Then one of the interesting problem would be to generalize Theorem 
\ref{thm-upper} for any integer $k \geq 2$, i.e., 
%the upper bounds for 
%m$ such that for any integer $k \geq 2$, i.e., 
%
\begin{itemize}
\item For any integer $k \geq 2$, 
show upper bounds for $m$ such that a random instance 
of the complete and strict $k$-weighted popular matching problem 
with $w_{i} \geq 2w_{i+1}$ $(\mbox{for each }1 \leq i \leq k-1)$ has a 
$k$-weighted popular matching with probability $1-o(1)$. 
\end{itemize}}

%OW: 新版
Theorem \ref{thm-lower} can be trivially generalized to any multiple category case;
that is, with the same proof, we have the following bound.
%
%In a way similar to the proof of Theorem \ref{thm-lower}, 
%we can show the following lower bounds for $m$ such that 
%for any integer $k\geq 2$, 
%a random instance of the complete and strict
%$k$-weighted matching problem does not have a $k$-weighted popular matching with 
%high probability, i.e.,  
%
\begin{thm} \label{thm-lower-k}
For any integer $k > 2$, if $m/n^{4/3}=o(1)$, then a 
random instance of the complete~and~strict $k$-weighted popular 
matching problem with $w_{i} \geq 2w_{i+1}$ $(1 \leq i \leq k-1)$ 
has a $k$-weighted popular matching with probability $o(1)$.
\end{thm}
Then an interesting problem is
to show some upper bound result by generalizing Theorem \ref{thm-upper}
for~any~integer $k>2$,
maybe under the condition that
$w_{i} \geq 2w_{i+1}$ for all $i$, $1 \leq i \leq k-1$.
%
%===========================

%====================
%
\newpage
\appendix
%
%=================================================================
\section{Proofs of Inequalities} \label{derivations-inequalities}
%=================================================================
%
%=====================================================
\subsection{Proof of Inequality (\ref{eq-2-common})} 
\label{derivations-inequality-2-common} 
%=====================================================
%
Let $\vec{v}=(x_{1},x_{2},y_{1},y_{2}) \in T$. 
For each $\vec{w}=(x_{1}',x_{2}',y_{1}',y_{2}') \in T_{2}(\vec{v})$, let us consider 
the following cases: (case-0) $\leng{\{y_{1},y_{2}\} \cap \{y_{1}',y_{2}'\}}=0$; 
(case-1) $\leng{\{y_{1},y_{2}\} \cap \{y_{1}',y_{2}'\}}=1$. 
Let 
\begin{eqnarray*}
T_{2}^{0}(\vec{v}) & = & 
\{\vec{w} \in T_{2}(\vec{v}): \leng{\{y_{1},y_{2}\} \cap \{y_{1}',y_{2}'\}}=0\};\\
T_{2}^{1}(\vec{v}) & = & 
\{\vec{w} \in T_{2}(\vec{v}): \leng{\{y_{1},y_{2}\} \cap \{y_{1}',y_{2}'\}}=1\}.
\end{eqnarray*}
For each $\vec{v} \in T$, it is immediate to see that 
$T_{2}^{0}(\vec{v}), T_{2}^{1}(\vec{v})$ is the 
partition of $T_{2}(\vec{v})$, and from the definitions of 
$T_{2}^{0}(\vec{v})$ and $T_{2}^{1}(\vec{v})$, we have 
that $\leng{T_{2}^{0}(\vec{v})}\leq n_{2}^{2}$; 
$\leng{T_{2}^{1}(\vec{v})}\leq 2n_{2}$. So 
from the assumption that $m-n\geq m/c$ for some 
constant $c > 1$, it follows that for each $\vec{v} \in T$, 
\begin{eqnarray}
\sum_{\vec{w} \in T_{2}^{0}(\vec{v})} {\bf E}\left[Z_{\vec{v}} Z_{\vec{w}}\right] 
& \leq & \sum_{\vec{w} \in T_{2}^{0}(\vec{v})} 
\frac{1}{m} \left(\frac{1}{m-n_{1}}\right)^{4} 
\leq \sum_{\vec{w} \in T_{2}^{0}(\vec{v})} 
\frac{1}{m} \left(\frac{1}{m-n}\right)^{4}\nonumber\\
& \leq & \sum_{\vec{w} \in T_{2}^{0}(\vec{v})} 
\frac{1}{m} \left(\frac{c}{m}\right)^{4}
= \frac{c^{4}}{m^{5}} \leng{T_{2}^{0}(\vec{v})} 
\leq  \frac{c^{4}}{m^{5}} n_{2}^{2}\nonumber\\
& = & \frac{c^{4}(1-\delta)^{2}}{m^{5}}n^{2}; \label{eq-T-2-0}\\
\sum_{\vec{w} \in T_{2}^{1}(\vec{v})} {\bf E}\left[Z_{\vec{v}} Z_{\vec{w}}\right] 
& \leq & \sum_{\vec{w} \in T_{2}^{1}(\vec{v})} 
\frac{1}{m} \left(\frac{1}{m-n_{1}}\right)^{3} 
\leq \sum_{\vec{w} \in T_{2}^{1}(\vec{v})} 
\frac{1}{m} \left(\frac{1}{m-n}\right)^{3}\nonumber\\
& \leq & \sum_{\vec{w} \in T_{2}^{1}(\vec{v})} 
\frac{1}{m} \left(\frac{c}{m}\right)^{3}
= \frac{c^{3}}{m^{4}} \leng{T_{2}^{1}(\vec{v})} 
\leq  \frac{2c^{3}}{m^{4}} n_{2}\nonumber\\
& = & \frac{2c^{3}(1-\delta)}{m^{4}}n. \label{eq-T-2-1}
\end{eqnarray}
Thus from Inequalities (\ref{eq-T-2-0}) and (\ref{eq-T-2-1}), we finally have that 
\begin{eqnarray*}
\sum_{\vec{v} \in T} 
\sum_{\vec{w} \in T_{2}(\vec{v})} 
{\bf E} \left[ Z_{\vec{v}} Z_{\vec{w}}\right] 
& = & \sum_{\vec{v} \in T} 
\sum_{\vec{w} \in T_{2}^{0}(\vec{v})} 
{\bf E} \left[ Z_{\vec{v}} Z_{\vec{w}}\right] + 
\sum_{\vec{v} \in T} 
\sum_{\vec{w} \in T_{2}^{1}(\vec{v})} 
{\bf E} \left[ Z_{\vec{v}} Z_{\vec{w}}\right]\\
& \leq & 
\sum_{\vec{v} \in T} \left\{
\frac{c^{4}(1-\delta)^{2}}{m^{5}} n^{2}
+ 
\frac{2c^{3}(1-\delta)}{m^{4}} n \right\}\\
& = & 
\left\{\frac{c^{4}(1-\delta)^{2}}{m^{5}} n^{2}
+ 
\frac{2c^{3}(1-\delta)}{m^{4}} n \right\} \leng{T}. 
\end{eqnarray*}
%
%====================================================
\subsection{Proof of Inequality (\ref{eq-1-common})} 
\label{derivations-inequality-1-common} 
%====================================================
%
Let $\vec{v}=(x_{1},x_{2},y_{1},y_{2}) \in T$. 
For each $\vec{w}=(x_{1}',x_{2}',y_{1}',y_{2}') \in T_{1}(\vec{v})$, we have 
the following cases: (case-0) $\leng{\{y_{1},y_{2}\}\cap \{y_{1}',y_{2}'\}}=0$; 
(case-1) $\leng{\{y_{1},y_{2}\}\cap \{y_{1}',y_{2}'\}}=1$; 
(case-2) $\leng{\{y_{1},y_{2}\}\cap \{y_{1}',y_{2}'\}}=2$. 
Let 
\begin{eqnarray*}
T_{1}^{0}(\vec{v}) & = & 
\{\vec{w} \in T_{1}(\vec{v}): \leng{\{y_{1},y_{2}\}\cap \{y_{1}',y_{2}'\}}=0\};\\
T_{1}^{1}(\vec{v}) & = & 
\{\vec{w} \in T_{1}(\vec{v}): \leng{\{y_{1},y_{2}\}\cap \{y_{1}',y_{2}'\}}=1\};\\
T_{1}^{2}(\vec{v}) & = & 
\{\vec{w} \in T_{1}(\vec{v}): \leng{\{y_{1},y_{2}\}\cap \{y_{1}',y_{2}'\}}=2\}. 
\end{eqnarray*}
For each $\vec{v} \in T$, it is immediate that 
$T_{1}^{0}(\vec{v}), T_{1}^{1}(\vec{v}), T_{1}^{2}(\vec{v})$
is the partition of $T_{1}(\vec{v})$, and from the definitions of 
$T_{1}^{0}(\vec{v})$, $T_{1}^{1}(\vec{v})$, and $T_{1}^{2}(\vec{v})$, 
we have that 
$\leng{T_{1}^{0}(\vec{v})}\leq 4n_{1}n_{2}^{2}$;  
$\leng{T_{1}^{1}(\vec{v})}\leq 4n_{1}n_{2}$; 
$\leng{T_{1}^{2}(\vec{v})}\leq 4n_{1}$. 
So from the assumption that $m-n\geq m/c$ for some 
constant $c > 1$, it follows that for each $\vec{v} \in T$, 
\begin{eqnarray}
\sum_{\vec{w} \in T_{1}^{0}(\vec{v})} 
{\bf E} \left[ Z_{\vec{v}} Z_{\vec{w}}\right] 
& \leq & \sum_{\vec{w} \in T_{1}^{0}(\vec{v})} 
\frac{1}{m^{2}} \left(\frac{1}{m-n_{1}}\right)^{4} 
\leq \sum_{\vec{w} \in T_{1}^{0}(\vec{v})} 
\frac{1}{m^{2}} \left(\frac{1}{m-n}\right)^{4}\nonumber\\
& \leq & \sum_{\vec{w} \in T_{1}^{0}(\vec{v})} 
\frac{1}{m^{2}} \left(\frac{c}{m}\right)^{4} 
= \frac{c^{4}}{m^{6}}\leng{T_{1}^{0}(\vec{v})} 
\leq \frac{4c^{4}}{m^{6}}n_{1}n_{2}^{2}\nonumber\\
& = & \frac{4c^{4}\delta(1-\delta)^{2}}{m^{6}}n^{3}; \label{eq-T-1-0}\\
\sum_{\vec{w} \in T_{1}^{1}(\vec{v})} 
{\bf E} \left[ Z_{\vec{v}} Z_{\vec{w}}\right] 
& \leq & \sum_{\vec{w} \in T_{1}^{1}(\vec{v})} 
\frac{1}{m^{2}} \left(\frac{1}{m-n_{1}}\right)^{3} 
\leq \sum_{\vec{w} \in T_{1}^{1}(\vec{v})} 
\frac{1}{m^{2}} \left(\frac{1}{m-n}\right)^{3}\nonumber\\
& \leq & \sum_{\vec{w} \in T_{1}^{1}(\vec{v})} 
\frac{1}{m^{2}} \left(\frac{c}{m}\right)^{3} 
= \frac{c^{3}}{m^{5}}\leng{T_{1}^{1}(\vec{v})} 
\leq \frac{4c^{3}}{m^{5}}n_{1}n_{2}\nonumber\\
& = & \frac{4c^{3}\delta(1-\delta)}{m^{5}}n^{2}; \label{eq-T-1-1}\\
\sum_{\vec{w} \in T_{1}^{2}(\vec{v})} 
{\bf E} \left[ Z_{\vec{v}} Z_{\vec{w}}\right] 
& \leq & \sum_{\vec{w} \in T_{1}^{2}(\vec{v})} 
\frac{1}{m^{2}} \left(\frac{1}{m-n_{1}}\right)^{3} 
\leq \sum_{\vec{w} \in T_{1}^{2}(\vec{v})} 
\frac{1}{m^{2}} \left(\frac{1}{m-n}\right)^{3}\nonumber\\
& \leq & \sum_{\vec{w} \in T_{1}^{2}(\vec{v})} 
\frac{1}{m^{2}} \left(\frac{c}{m}\right)^{3} 
= \frac{c^{3}}{m^{5}}\leng{T_{1}^{2}(\vec{v})} 
\leq \frac{4c^{3}}{m^{5}}n_{1}\nonumber\\
& = & \frac{4c^{3}\delta}{m^{5}}n. \label{eq-T-1-2}
\end{eqnarray}
Thus from Inequalities (\ref{eq-T-1-0}), (\ref{eq-T-1-1}), and 
(\ref{eq-T-1-2}), we finally have that 
\begin{eqnarray*}
\sum_{\vec{v} \in T} 
\sum_{\vec{w} \in T_{1}(\vec{v})} 
{\bf E} \left[ Z_{\vec{v}} Z_{\vec{w}}\right] 
& = & \sum_{\vec{v} \in T} 
\sum_{\vec{w} \in T_{1}^{0}(\vec{v})} 
{\bf E} \left[ Z_{\vec{v}} Z_{\vec{w}}\right] + 
\sum_{\vec{v} \in T} 
\sum_{\vec{w} \in T_{1}^{1}(\vec{v})} 
{\bf E} \left[ Z_{\vec{v}} Z_{\vec{w}}\right] + 
\sum_{\vec{v} \in T} 
\sum_{\vec{w} \in T_{1}^{2}(\vec{v})} 
{\bf E} \left[ Z_{\vec{v}} Z_{\vec{w}}\right]\\
& \leq & 
\sum_{\vec{v} \in T} 
\left\{
\frac{4c^{4}\delta(1-\delta)^{2}}{m^{6}}n^{3}+ 
\frac{4c^{3}\delta(1-\delta)}{m^{5}}n^{2}+ 
\frac{4c^{3}\delta}{m^{5}}n \right\}\\
& = & \left\{
\frac{4c^{4}\delta(1-\delta)^{2}}{m^{6}}n^{3}+ 
\frac{4c^{3}\delta(1-\delta)}{m^{5}}n^{2}+ 
\frac{4c^{3}\delta}{m^{5}}n \right\} \leng{T}.
\end{eqnarray*}
%
%====================================================
\subsection{Proof of Inequality (\ref{eq-0-common})} 
\label{derivations-inequality-0-common} 
%====================================================
%
Let $\vec{v}=(x_{1},x_{2},y_{1},y_{2}) \in T$. 
For each $\vec{w}=(x_{1}',x_{2}',y_{1}',y_{2}') \in T_{0}(\vec{v})$, we have 
the following cases: (case-0) $\leng{\{y_{1},y_{2}\}\cap \{y_{1}',y_{2}'\}}=0$; 
(case-1) $\leng{\{y_{1},y_{2}\}\cap \{y_{1}',y_{2}'\}}=1$; 
(case-2) $\leng{\{y_{1},y_{2}\}\cap \{y_{1}',y_{2}'\}}=2$. 
Let 
\begin{eqnarray*}
T_{0}^{0}(\vec{v}) & = & 
\{\vec{w} \in T_{0}(\vec{v}): \leng{\{y_{1},y_{2}\}\cap \{y_{1}',y_{2}'\}}=0\};\\
T_{0}^{1}(\vec{v}) & = & 
\{\vec{w} \in T_{0}(\vec{v}): \leng{\{y_{1},y_{2}\}\cap \{y_{1}',y_{2}'\}}=1\};\\
T_{0}^{2}(\vec{v}) & = & 
\{\vec{w} \in T_{0}(\vec{v}): \leng{\{y_{1},y_{2}\}\cap \{y_{1}',y_{2}'\}}=2\}. 
\end{eqnarray*}
For each $\vec{v} \in T$, it is immediate that 
$T_{0}^{0}(\vec{v}), T_{0}^{1}(\vec{v}), T_{0}^{2}(\vec{v})$
is the partition of $T_{0}(\vec{v})$. For any 
$\vec{w} \in T_{0}^{0}(\vec{v})$, it is obvious that 
$\Pr[Z_{\vec{v}}=1\wedge Z_{\vec{w}}=1]=\Pr[Z_{\vec{v}}=1]\times 
\Pr[Z_{\vec{w}}=1]$, which implies that
\begin{eqnarray}
\sum_{\vec{v} \in T} \sum_{\vec{w} \in T_{0}^{0}(\vec{v})} 
{\bf E}[Z_{\vec{v}}Z_{\vec{w}}] & = & 
\sum_{\vec{v} \in T} \sum_{\vec{w} \in T_{0}^{0}(\vec{v})} 
\Pr[Z_{\vec{v}}=1 \wedge Z_{\vec{w}}=1]\nonumber\\
& = & \sum_{\vec{v} \in T} \sum_{\vec{w} \in T_{0}^{0}(\vec{v})} 
\Pr[Z_{\vec{v}}=1]\times \Pr[Z_{\vec{w}}=1]\nonumber\\
& = & \sum_{\vec{v} \in T} \Pr[Z_{\vec{v}}=1] \sum_{\vec{w} \in T_{0}^{0}(\vec{v})} 
\Pr[Z_{\vec{w}}=1]\nonumber\\
& \leq & \sum_{\vec{v} \in T} \Pr[Z_{\vec{v}}=1] \sum_{\vec{w} \in T} 
\Pr[Z_{\vec{w}}=1]\nonumber\\
& = & {\bf E}^{2}[Z]. \label{eq-T-0-0}
\end{eqnarray}
From the definitions of $T_{0}^{1}(\vec{v})$ and $T_{0}^{2}(\vec{v})$, we have that 
$\leng{T_{0}^{1}(\vec{v})} \leq 2n_{1}^{2}n_{2}$; 
$\leng{T_{0}^{2}(\vec{v})} \leq n_{1}^{2}$. Then from~the~as\-sumption that 
$m-n\geq m/c$ for some constant $c>1$, it follows that for each $\vec{v} \in T$, 
\begin{eqnarray}
\sum_{\vec{w} \in T_{0}^{1}(\vec{v})} {\bf E}[Z_{\vec{v}}Z_{\vec{w}}] & = & 
\sum_{\vec{w} \in T_{0}^{1}(\vec{v})} \frac{1}{m^{2}}
\left( \frac{1}{m - n_{1}}\right)^{4}
\leq \sum_{\vec{w} \in T_{0}^{1}(\vec{v})} \frac{1}{m^{2}}
\left( \frac{1}{m - n}\right)^{4}\nonumber\\
& \leq & \sum_{\vec{w} \in T_{0}^{1}(\vec{v})} \frac{1}{m^{2}}
\left( \frac{c}{m}\right)^{4} = \frac{c^{4}}{m^{6}}\leng{T_{0}^{1}(\vec{v})}
\leq \frac{2c^{4}}{m^{6}}n_{1}^{2}n_{2}\nonumber\\
& = & \frac{2c^{4}\delta^{2}(1-\delta)}{m^{6}}n^{3}; \label{eq-T-0-1}\\
\sum_{\vec{w} \in T_{0}^{2}(\vec{v})} {\bf E}[Z_{\vec{v}}Z_{\vec{w}}] & = & 
\sum_{\vec{w} \in T_{0}^{2}(\vec{v})} \frac{1}{m^{2}}
\left( \frac{1}{m - n_{1}}\right)^{4}
\leq \sum_{\vec{w} \in T_{0}^{2}(\vec{v})} \frac{1}{m^{2}}
\left( \frac{1}{m - n}\right)^{4}\nonumber\\
& \leq & \sum_{\vec{w} \in T_{0}^{2}(\vec{v})} \frac{1}{m^{2}}
\left( \frac{c}{m}\right)^{4} = \frac{c^{4}}{m^{6}}\leng{T_{0}^{2}(\vec{v})}
\leq \frac{c^{4}}{m^{6}}n_{1}^{2}\nonumber\\
& = & \frac{c^{4}\delta^{2}}{m^{6}}n^{2}. \label{eq-T-0-2}
\end{eqnarray}
Thus from Inequalities (\ref{eq-T-0-0}), (\ref{eq-T-0-1}), and (\ref{eq-T-0-2}), 
we finally have that 
\begin{eqnarray*}
\sum_{\vec{v} \in T} \sum_{\vec{w} \in T_{0}(\vec{w})} 
{\bf E}[Z_{\vec{v}}Z_{\vec{w}}] & = & 
\sum_{\vec{v} \in T} \sum_{\vec{w} \in T_{0}^{0}(\vec{w})} 
{\bf E}[Z_{\vec{v}}Z_{\vec{w}}] + 
\sum_{\vec{v} \in T} \sum_{\vec{w} \in T_{0}^{1}(\vec{w})} 
{\bf E}[Z_{\vec{v}}Z_{\vec{w}}] + 
\sum_{\vec{v} \in T} \sum_{\vec{w} \in T_{0}^{2}(\vec{w})} 
{\bf E}[Z_{\vec{v}}Z_{\vec{w}}]\\
& \leq & {\bf E}^{2}[Z] + \sum_{\vec{v} \in T} 
\left\{ \frac{2c^{4}\delta^{2}(1-\delta)}{m^{6}}n^{3} + 
\frac{c^{4}\delta^{2}}{m^{6}}n^{2} \right\}\\
& = & {\bf E}^{2}[Z] + 
\left\{ \frac{2c^{4}\delta^{2}(1-\delta)}{m^{6}}n^{3} + 
\frac{c^{4}\delta^{2}}{m^{6}}n^{2} \right\}\leng{T}. 
\end{eqnarray*}
\end{document}